\documentclass[aps,prl,twocolumn,longbibliography,superscriptaddress]{revtex4-2}

\usepackage{amssymb}
\usepackage{amsbsy}
\usepackage{amsmath}
\usepackage{graphicx}
\usepackage{graphics}
\usepackage{setspace}
\usepackage{array}
\usepackage{color}
\usepackage{fontenc}
\usepackage{textcomp}
\usepackage{bm}
\usepackage{float}
\usepackage{pifont}
\usepackage[bookmarks=false,linkcolor=blue,urlcolor=blue,colorlinks,citecolor=blue]{hyperref}
\usepackage{soul}
\usepackage[T1]{fontenc}
\newcommand{\cmark}{\ding{51}}%
\newcommand{\xmark}{\ding{55}}%

\newcommand{\blue}[1]{{\color{blue}{#1}}}

\newcommand{\bsub}{\begin{subequations}}
\newcommand{\esub}{\end{subequations}}

\newcommand{\tr}{\text{tr}}

\graphicspath{{../Figures/}}

\definecolor{amethyst}{rgb}{0.6, 0.4, 0.8}

\begin{document}
\title{Quantum geometry induced nonlinear transport in altermagnets}
\author{Yuan Fang}
\affiliation{Department of Physics and Astronomy, Stony Brook University, Stony Brook, New York 11794, USA}
\affiliation{Department of Physics and Astronomy, Rice Center for Quantum Materials, Rice University, Houston, Texas 77005, USA}

\author{Jennifer Cano}
\affiliation{Department of Physics and Astronomy, Stony Brook University, Stony Brook, New York 11794, USA}
\affiliation{Center for Computational Quantum Physics, Flatiron Institute, New York, New York 10010, USA}

\author{Sayed Ali Akbar Ghorashi}\email[Correspondence\,to:\,]{sayedaliakbar.ghorashi@stonybrook.edu}
\affiliation{Department of Physics and Astronomy, Stony Brook University, Stony Brook, New York 11794, USA}

\date{\today}

\newcommand{\be}{\begin{equation}}
\newcommand{\ee}{\end{equation}}
\newcommand{\bea}{\begin{eqnarray}}
\newcommand{\eea}{\end{eqnarray}}
\newcommand{\h}{\hspace{0.30 cm}}
\newcommand{\vs}{\vspace{0.30 cm}}
\newcommand{\n}{\nonumber}


\begin{abstract}
Quantum geometry plays a pivotal role in the second-order response of $\cal PT$-symmetric antiferromagnets. Here we study the nonlinear response of 2D altermagnets protected by $C_n\cal T$ symmetry and show that their leading nonlinear response is third-order. Furthermore, we show that the contributions from the quantum metric and Berry curvature enter separately: the longitudinal response for all planar altermagnets \emph{only} has a contribution from the quantum metric quadrupole (QMQ), while transverse responses in general have contributions from both the Berry curvature quadrupole (BCQ) and QMQ. We show that for the well-known example of $d$-wave altermagnets the Hall response is dominated by the BCQ. Both longitudinal and transverse responses are strongly dependent on the crystalline anisotropy. While altermagnets are strictly defined in the limit of vanishing SOC, real altermagnets exhibit weak SOC, which is essential to observe this response. Specifically, SOC gaps the spin-group protected nodal line, generating a response peak that is sharpest when SOC is weak. 
Two Dirac nodes also contribute a divergence to the nonlinear response, whose scaling changes as a function of SOC. Finally, we apply our results to thin films of the 3D altermagnet RuO$_2$. 
Our work uncovers distinct features of altermagnets in nonlinear transport, providing experimental signatures as well as a guide to disentangling the different components of their quantum geometry.
\end{abstract}

\maketitle

\blue{\emph{Introduction}}.---Altermagnets are collinear antiferromagnets with weak spin-orbit coupling (SOC) that share properties of both ferromagnets and antiferromagnets, most prominently, spin-splitting in momentum space, while maintaining zero net magnetization \cite{altermagnet1, altermagnet2}. Unlike conventional antiferromagnets in which two magnetic sublattices are related by translation or inversion, in altermagnets they are related by a spin-group element, consisting of an independent spin flip and $C_n$ spatial rotation. However, this is strictly true only for vanishing SOC: the spin group reduces to a $C_n \mathcal{T}$ magnetic group when SOC is present \cite{altermagnet1,PhysRevX.12.021016,xiao2023spin,chen2023spin,yang2023symmetry,jiang2023enumeration}. Moreover, weak SOC is crucial in observing various phenomena measured in altermagnets, such as the anomalous Hall effect \cite{feng2022anomalous,ghorashi2023altermagnetic,fernandes2023topological}.     
\\
\indent Recently, there has been a surge of interest in the nonlinear response of topological $\mathcal{PT}$-symmetric antiferromagnets (AFMs), which is dominated by the nontrivial quantum geometry \cite{PhysRevLett.127.277201,PhysRevLett.127.277202,kaplan2023unifying,kaplan2022unification, QMDscience,QMDnature}. Unlike the time-reversal invariant nonlinear response in non-centrosymmetric materials \cite{Fu2015BCD,ma2019observation,du2021nonlinearReview,lai2021third}, $\mathcal{PT}$ symmetry requires the Berry curvature dipole (BCD) vanish and the nonlinear response is instead dominated by the scattering time $\tau$ independent quantum metric dipole (QMD).
\begin{figure}[ht]
    \centering
    \includegraphics[width=0.49\textwidth]{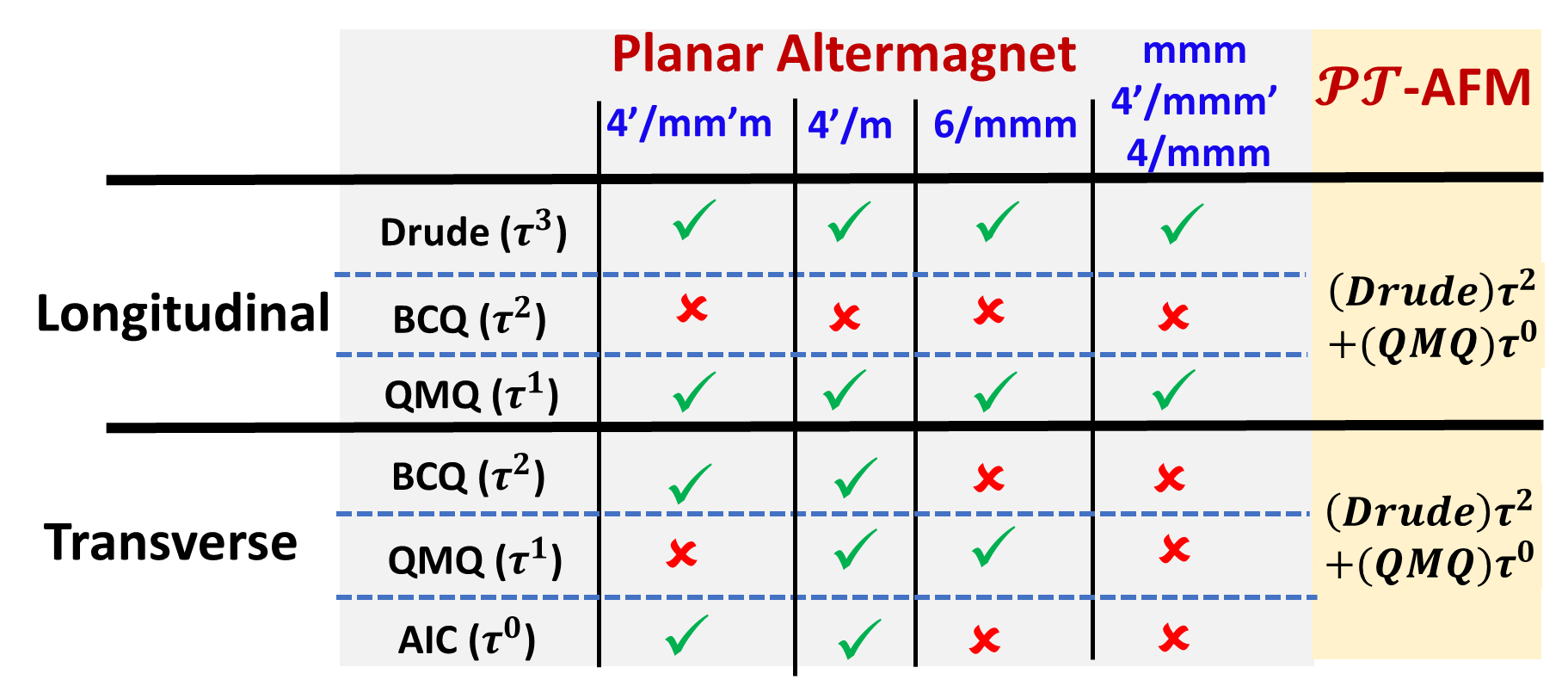}
    \caption{Third-order response in planar altermagnets protected by $C_n \cal T$ compared to $\mathcal{PT}$-AFMs. Unlike $\mathcal{PT}$-AFMs, the quantum metric and Berry curvature play distinct roles: only the QMQ enters the longitudinal response, while the Hall responses
generically have contributions from both the QMQ and BCQ, which can be disentangled by their scaling with $\tau$.}
    \label{fig:Adfig}
\end{figure}
\\
\indent Motivated by these developments, in this work we investigate the nonlinear responses in altermagnets, unravelling the role of quantum geometry order by order in $\tau$, as summarized in Fig.~\ref{fig:Adfig}.
Our derivation of nonlinear conductivity holds for any magnet with $C_n\mathcal{T}$ symmetry.
However, we find
distinct features in the nonlinear response of altermagnets that become sharper upon approaching the ideal altermagnet with vanishing SOC. \\
\indent 
Specifically, the leading nonlinear response in altermagnets -- or any magnet with $C_n\mathcal{T}$ symmetry -- is third-order. Similar to the linear anomalous Hall conductivity (AHC) in these materials, it exhibits crystalline anisotropy.
Moreover, unlike the $\mathcal{PT}$-symmetric AFMs, where both the longitudinal and Hall responses are dominated by the QMD, in altermagnets and other magnets preserving $C_n\mathcal{T}$ symmetry, the quantum geometric contribution to the longitudinal response is only from the quantum metric quadrupole (QMQ), while the Hall responses generically have contributions from both the QMQ and Berry curvature quadrupole (BCQ), which can be disentangled by how they scale with $\tau$. Further, in certain symmetry groups, the contribution from the QMQ, BCQ, or both, may vanish.
For example, due to the $d$-wave symmetry describing RuO$_2$, the Hall response only has a BCQ contribution; thus the longitudinal and transverse responses separately measure the QMQ and BCQ.
\\
\indent
The response of altermagnets with weak SOC differs from larger SOC magnets by the magnitude of these effects. Specifically, altermagnets exhibit
(i) a strong corrections from the subleading terms near Dirac nodes and (ii) 
a peak due to nodal line anti-crossings,
which is suppressed for large SOC.  
To demonstrate the experimental relevance of our results, we find a large nonlinear current for a thin film of RuO$_2$.\\
\indent \blue{\emph{Third-order response}}.--- We start by sketching the derivation of the third-order conductivity in an electric field, $\mathbf{E}$, which takes the form: $j_d^{(3)}=\sigma^{abc;d}E_{a}E_bE_c$.
The quantum metric and Berry curvature will play a pivotal role;  they are defined as the real symmetric and imaginary anti-symmetric parts of the quantum geometric tensor,
${\mathcal{G}}_{a b} = \tr(P\partial_a P\partial_b P) = {\mathfrak{g}}_{a b} - \frac i2 \Omega_{a b}$, where $P$ is a projector onto bands of interest and the trace is over internal indices.
\\
\indent The current is given by $\mathbf{j}=-e\int_k \mathbf{v}f$, where $\mathbf{v}$ is the group velocity, $f$ is the electron density and $e=|e|$ is the positive electron charge.
To proceed we adopt the semiclassical theory of wavepackets, starting with the Boltzmann equation in the relaxation time approximation,
\begin{align}
    \partial_t f+\frac{\mathbf{F}}{\hbar}  \cdot \nabla_{\mathbf{k}} f &=-\frac{f-f_0}{\tau},
\end{align}
where $\mathbf{F}=-\,e\mathbf{E}$ for electric field $\mathbf{E}$, $f_0$ is the Fermi-Dirac distribution, and $\tau$ is the scattering time. We solve for $f$ in the frequency domain by expanding in powers of $\mathbf E$. The $l$-th order perturbation to the charge density is
\begin{equation}
    f_n^{(l)} = \left(\frac{ e/{\hbar}}{i\omega + 1/\tau} \right)^l{\mathbf E}^l \nabla_{\mathbf{k}}^l f_n^{(0)} \xrightarrow{\omega \rightarrow 0} \left( \frac{e\tau}{\hbar} \right)^l{\mathbf E}^l \nabla_{\mathbf{k}}^l f_n^{(0)},
\end{equation}
where $n$ is band index.
The group velocity is determined semiclassically by 
\begin{equation}
    {\mathbf v}_n = \frac{1}{\hbar}\frac{\partial \varepsilon_n}{\partial {\mathbf k}} - \frac{e}{\hbar} {\mathbf E} \times {\mathbf \Omega_n},
\end{equation}
where $\mathbf \Omega_n$ is the Berry curvature of the $n$-th band.
\\
\indent Equating all the terms with the same order in $\mathbf{E}$ yields the nonlinear responses. We obtain the third-order conductivity (see \cite{sm} for details) 
\begin{align}\label{3rdS}
     &\sigma^{abc;d}= \tau^{3}\left[\frac{e^4}{\hbar^4} \sum_n \int_{\boldsymbol{k}} f_n \partial_{k^a} \partial_{k^b} \partial_{k^c} \partial_{k^d} \varepsilon_n\right] \cr
    -&\tau^2\left[\frac{e^4}{\hbar^3} \sum_n \int_{\boldsymbol{k}} f_n \frac13 \left(\partial_{k^a} \partial_{k^b} \Omega_n^{c d}+\partial_{k^b} \partial_{k^c} \Omega_n^{a d}+\partial_{k^a} \partial_{k^c} \Omega_n^{b d}\right) \right]\cr
    +&\tau\Bigg[\frac{e^4}{\hbar^2} \sum_n \int_{\boldsymbol{k}} f_n \frac13 \Big(2 ( \partial_{k^a} \partial_{k^d} G_n^{bc}+\partial_{k^b} \partial_{k^d} G_n^{ac}+\partial_{k^c} \partial_{k^d} G_n^{ab} ) \cr
    &\quad-\left(\partial_{k^a}\partial_{k^c} G_n^{b d}+\partial_{k^b}\partial_{k^c} G_n^{a d}+\partial_{k^a}\partial_{k^b} G_n^{cd} \right) \Big)\Bigg] \cr
    & +\tau^0 \text{AIC}
\end{align}
where $G^{ab}_n=\sum_{m\neq n} \frac{A^a_{nm}A^b_{mn}+A^b_{nm}A^a_{mn}}{2\varepsilon_{nm}}$ is the band-normalized quantum metric, $A_{nm}=\langle n|\mathbf r|m\rangle$ and $\varepsilon_{nm}=\varepsilon_n-\varepsilon_m$.
The first, second and third terms in Eq.~(\ref{3rdS}) are proportional to the third-order Drude, BCQ, and band-renormalized QMQ, respectively, which are each accompanied by a different power  of  $\tau$. The additional inter-band contribution (AIC) term takes the following form in a two band approximation (see \cite{sm} for its general form)
\begin{equation}
    \text{AIC} = -\frac{e^4}{\hbar}\sum_n \int_{\boldsymbol k}f_n \frac{2}{3\varepsilon_{n\bar n}} \left( G^{ab}_n \Omega^{cd}_n + G^{ac}_n \Omega^{bd}_n + G^{bc}_n \Omega^{ad}_n \right)
\end{equation}
where $\bar n \neq n$. The AIC term contains products of the Berry curvature and the quantum metric.
In altermagnets this term contributes near Dirac points but not around the anti-crossings that we will describe.
\\
\indent \blue{\emph{Nonlinear response of altermagnets}}.--- 
We now discuss the effect of $C_n\mathcal{T}$ symmetry on each term in Eq.~\eqref{3rdS}. 
We are interested in quasi-2D systems where the Neel vector, $\mathcal{N}$, is out of plane and the $\mathbf{k}$-dependence of the magnetization is in-plane.
\\
\indent The quantum metric $\mathfrak{g}_{a b} $ transforms as a rank-2 symmetric tensor under spatial symmetries while Berry curvature transforms as a rank-2 antisymmetric tensor which is dual to a pseudo-vector $\Omega^{c}\equiv\epsilon^{abc} \Omega_{a b} $ in 3D. Under a spatial symmetry $g$, momentum $\boldsymbol{k}$ maps to $R_g\boldsymbol{k}$ where $R_g$ is the rotational part of the symmetry. The quantum metric and Berry curvature transform as, 
\begin{align}
\label{eqn:QMunderg}
  \mathfrak{g}_{a' b'}(R_{g}\boldsymbol{k}) = & R_g{}_{a'}^{a}R_g{}_{b'}^{b} \mathfrak{g}_{a b}(\boldsymbol{k}),\\
  \Omega^{c'}(R_{g}\boldsymbol{k}) = & (-)^{\chi(g)} \det(R_g) ~ R_{g}{}^{c'}_{c} \Omega^{c}(\boldsymbol{k})
\end{align} 
where $\chi(g)=\pm 1$ for a unitary/antiunitary symmetry $g$ and $\det R_g=\pm 1$ for proper/improper rotation. 
The $\chi(g)$ term in the Berry curvature transformation comes from the Berry curvature being the imaginary part of the quantum geometric tensor. 
The symmetry transformations are derived in~\cite{sm}.\\
\indent Spin group symmetries take the form $[g_s||R_g, \mathbf{t}]$, where $\mathbf{t}$ is a (sub)lattice translation. Here, $g_s$ acts in spin space while $R_g$ and $\mathbf{t}$ act on real space.
Time-reversal symmetry simultaneously acts on both spaces as, ${\cal T} = [C_{2s}||E]{\cal K}$ where $C_{2s}$ indicates a two-fold rotation in spin space and ${\cal K}$ is the complex conjugation operator. 
In general, spin groups contain elements that are products of time-reversal and unitary symmetries.
Since the spin operations and pure translations do not constrain the quantum metric or Berry curvature \cite{sm}, the symmetry constraints of the spin group are identical to those of the magnetic point group formed by the spatial part $[E_s||R_{g}{}]$, where $E_s$ denotes identity in spin space. 
Recently, Ref.~\cite{LawPRB2023} classified Berry curvature multipoles subject to magnetic point group symmetries. In this Letter, we present an algorithm to classify the quantum metric multipoles. The tables of allowed terms for point groups corresponding to all the planar $(d,g,i)$-wave altermagnets are shown in \cite{sm}.

Henceforth, we focus on planar $d$-wave altermagnets,
such as RuO$_2$, MnF$_2$ and Mn$_5$Si$_3$ \cite{altermagnet1,reichlova2021macroscopic}. We present a similar symmetry analysis in \cite{sm} for $g$- and $i$-wave models, relevant to materials such as KMnF$_2$;
more materials may be realized by applied field or chemical substitution, as proposed in
\cite{mazin2023induced}.\\
\indent
We now specialize to a
$d$-wave altermagnet invariant under the spin group $^24/^1m{}^1m{}^2m$ (without SOC) and magnetic group $4'/mm'm$ (with SOC).
We start with constraints from the magnetic point group $4'/mm'm$ symmetries. 
For the first order response, the longitudinal Drude term $\partial_{k^x}\partial_{k^x}\epsilon$ is the only term that survives.
This is because due to $C_4{\cal T}$, the integral of Berry curvature and the transverse Drude term $\partial_{k^x}\partial_{k^y}\epsilon$ vanish; thus, all the contributions to the first-order transverse conductivity diminish altogether. 
The second-order responses are forbidden by inversion symmetry~\cite{sm}.
Therefore, the leading nonlinear conductivity in altermagnets is the third-order response. 
The following components of the QMQ and BCQ are symmetry-allowed: $\partial_{k^x}\partial_{k^x}\mathfrak{g}_{xx}$, $\partial_{k^x}\partial_{k^x}\mathfrak{g}_{yy}$, $\partial_{k^x}\partial_{k^y}\mathfrak{g}_{xy}$, $\partial_{k^x}\partial_{k^x}\Omega_{xy}$, $\partial_{k^y}\partial_{k^y}\Omega_{xy}$ and $\partial_{k^x}\partial_{k^y}\Omega_{xy}$. Similarly, the third-order Drude terms with even numbers of $\partial_{k^x}$ or $\partial_{k^y}$ are allowed. 

While symmetry identifies which components of conductivity are in general non-vanishing, it does not specify how each term contributes.
In altermagnets, the main contributions to the third-order response are from two features in the band structure: (i) Dirac nodes and (ii) nodal line anti-crossings.
The latter is specific to altermagnets and warrants some explanation:
in the absence of SOC, the spin group enforces nodal lines in the band structure. However, in real altermagnets, weak SOC is present.
SOC gaps the nodal line, resulting in an anti-crossing that gives rise to a pronounced peak in the third order conductivity. 
The peak broadens as SOC is further increased; thus it is a specific feature of altermagnets that will be weak or absent for other materials in the same magnetic group, but with strong SOC.
The response of altermagnets is further distinguished from magnets with strong SOC by strong corrections from the subleading terms near Dirac points. 
%
\begin{figure}[t]
    \centering
    \includegraphics[width=\linewidth]{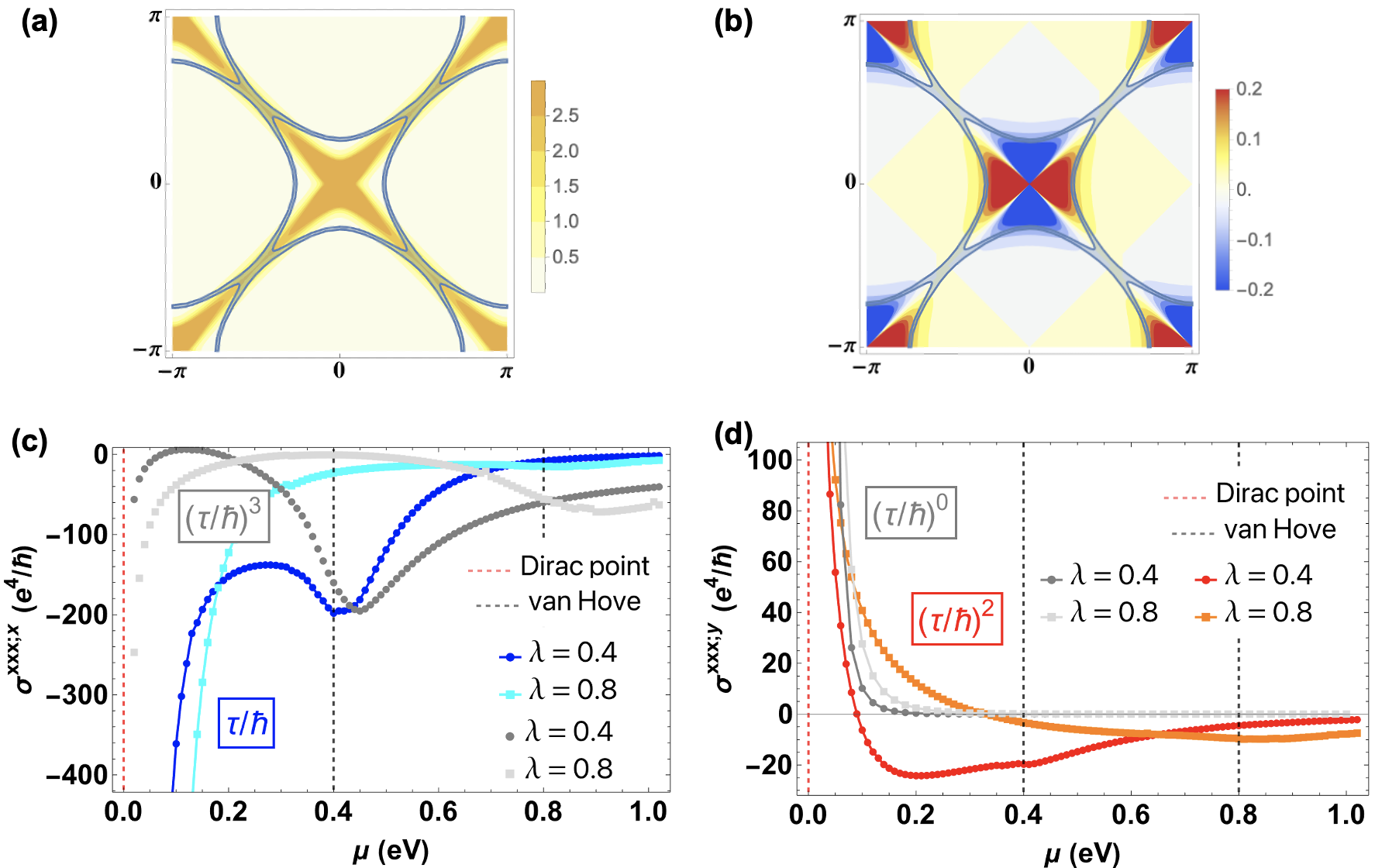}
    \caption{(a) Quantum metric $\mathfrak{g}_{xx}$
    and (b) Berry curvature overlaid by the Fermi surface at $\mu=\lambda$, where the van Hove singularity appears along the anti-crossing lines $k_x=\pm k_y$.
    (c) $\sigma^{xxx;x}$ and (d) $\sigma^{xxx;y}$ for two values of SOC, $\lambda$.
    The Drude $(\tau^3)$, QMQ  $(\tau)$, BCQ  $(\tau^2)$ and AIC  $(\tau^0)$ contributions are shown separately, indicated by their $\tau$ dependence. 
    The peak near the van Hove energy is visible in the Drude and QMQ terms.
    The parameters used are $J= 1$eV, $\lambda=0.4,0.8$eV.}
    \label{fig:Fig2}
\end{figure}
\\
\indent We explicitly demonstrate these results using an effective two-band model, which characterizes the physics of quasi-2D planar $d_{x^2-y^2}$-wave altermagnets \cite{altermagnet1}. 
The two-band model is defined on a square lattice with the magnetic atoms sitting at sites $A=(\frac12,0)$ and $B=(0,\frac12)$. 
In the altermagnetic phase, the magnetic moments are up/down on the $A/B$ sites.
The Hamiltonian is
\begin{multline}
\label{eqn:Ham_Toy}
    H= J (\cos k_x -\cos k_y)\sigma_z \\ +\lambda\left[\sin(\frac{k_x+k_y}{2})\sigma_x+\sin(\frac{k_y-k_x}{2})\sigma_y \right],
\end{multline}
where $J$ denotes the $d_{x^2-y^2}$ altermagnetic order parameter, $\lambda$ incorporates SOC,
and the Pauli matrices act in spin space. 
Eq.~\eqref{eqn:Ham_Toy} possesses two gapless Dirac nodes at $\Gamma = (0,0)$ and $M=(\pi,\pi)$. 
Moreover, Eq.~\eqref{eqn:Ham_Toy} hosts van-Hove singularities at momenta $|k_x| = |k_y| = \pi/2$ and energies $\pm \lambda$.
As we show below the anti-crossings that occur around these van-Hove points significantly affect the nonlinear response of altermagnets. 

The quantum metric, $\mathfrak{g}_{xx}$, and Berry curvature distribution in the BZ are overlaid with the Fermi surface at the energy of the van Hove singularity, $\mu=\lambda$, in Fig.~\ref{fig:Fig2}(a) and (b) respectively. 
Both quantities are peaked near the anti-crossing lines $k_x = \pm k_y$.

We now show how the QMQ and BCQ enter different components of the conductivity tensor. Our symmetry analysis reveals only four independent components of the nonlinear conductivity. We here discuss the two components most commonly measured, the longitudinal, $\sigma^{xxx;x}$, and transverse, $\sigma^{xxx;y}$, conductivity, which for a $d_{x^2-y^2}$-wave altermagnet takes the form, 
\begin{align}\label{Res1}
       \sigma^{xxx;x} & = \tau^3 \frac{e^4}{\hbar^4}\int\partial_x\partial_x\partial_x\partial_x\varepsilon_n + \tau \frac{e^4}{\hbar^2}\int \partial_x\partial_x G^{xx}_n \cr
    \sigma^{xxx;y} & = -\tau^2 \frac{e^4}{\hbar^3} \int \partial_x\partial_x\Omega^{xy}_n - \frac{e^4}{\hbar}\int \frac{2G^{xx}_n\Omega^{xy}_n}{\varepsilon_{n\bar n}}
\end{align}
where the band indices and summation over filled bands are implicit and the two-band limit is assumed for the second term (AIC term) in $\sigma^{xxx;y}$.
For $\mathcal{PT}$-AFMs, due to the vanishing Berry curvature, both the longitudinal and Hall responses have the same quantum geometric origin i.e, QMD. In contrast, for $d$-wave altermagnets in magnetic group $4'/mm'm$, the QMQ and BCQ both appear and play distinct roles:  only the QMQ enters the longitudinal response, while only the BCQ enters the transverse response.
Therefore, these altermagnets separately probe each component of the quantum geometry. For other planar altermagnets, the transverse response may have contributions from both the BCQ and QMQ, but for all the planer altermagnets the longitudinal component only receives a  contribution from the QMQ.
Notice that each term in Eq~(\ref{Res1}) 
scales differently with $\tau$, allowing the quantum geometric contributions to be extracted experimentally.
\\
\indent Fig.~\ref{fig:Fig2}(c,d) show contributions to $\sigma^{xxx;x}$ and $\sigma^{xxx;y}$ as a function of chemical potential for different strengths of SOC. There are two main contributions to the third-order response, one originating from the divergence of the quantum geometry and Drude terms around the Dirac nodes and a peak around the anti-crossing that results from SOC gapping the spin-group protected nodal lines along $k_x=\pm k_y$. For larger values of $\lambda$, the anti-crossing peak diminishes. It is optimized for weak SOC, which is precisely the regime of real altermagnets.
As mentioned above, Fig.~\ref{fig:Fig2}(d) demonstrates that the AIC term only contributes when the chemical potential is near the Dirac node. Therefore, near the anti-crossing the third-order Hall response is dominated by the BCQ. 
\\
\indent The longitudinal and transverse nonlinear conductivities in altermagnets exhibit different behaviour near the Dirac nodes compared to magnets with larger SOC. 
As the chemical potential approaches the Dirac point, i.e.,  $\mu \ll \lambda^2/ J$, the integrals entering the conductivities take the form
\begin{align}
\label{div}
    \int_k \partial_x\partial_x\Omega^{xy} &\rightarrow  -\frac{\pi J}{8|\mu|} f(\Lambda),\quad \text{AIC} \rightarrow -\frac{\pi J}{384 |\mu|^3}f(\Lambda)\cr
    \int_k \partial_x\partial_xG^{xx} &\rightarrow -\frac{5\pi \lambda^2}{64|\mu|^3} f(\Lambda)\,
    \int_k \partial_x\partial_x\partial_x\partial_x \epsilon \rightarrow -\frac{3\pi \lambda^2}{8|\mu|} f(\Lambda)
\end{align}
where $f(\Lambda)=1+O\big(\Lambda\big),\,\Lambda=\mu^2J^2/\lambda^4$.  By comparing to Eq.~\eqref{Res1}, Eq.~\eqref{div} explains the opposite signs of the longitudinal and Hall responses as they approach the Dirac node. 
In altermagnets where $\lambda \ll J$, the leading $1/|\mu|^3$ 
term in both longitudinal and transverse responses may receive strong corrections from the subleading terms.
\\
\begin{figure}[t]
    \centering
    \includegraphics[width=\linewidth]{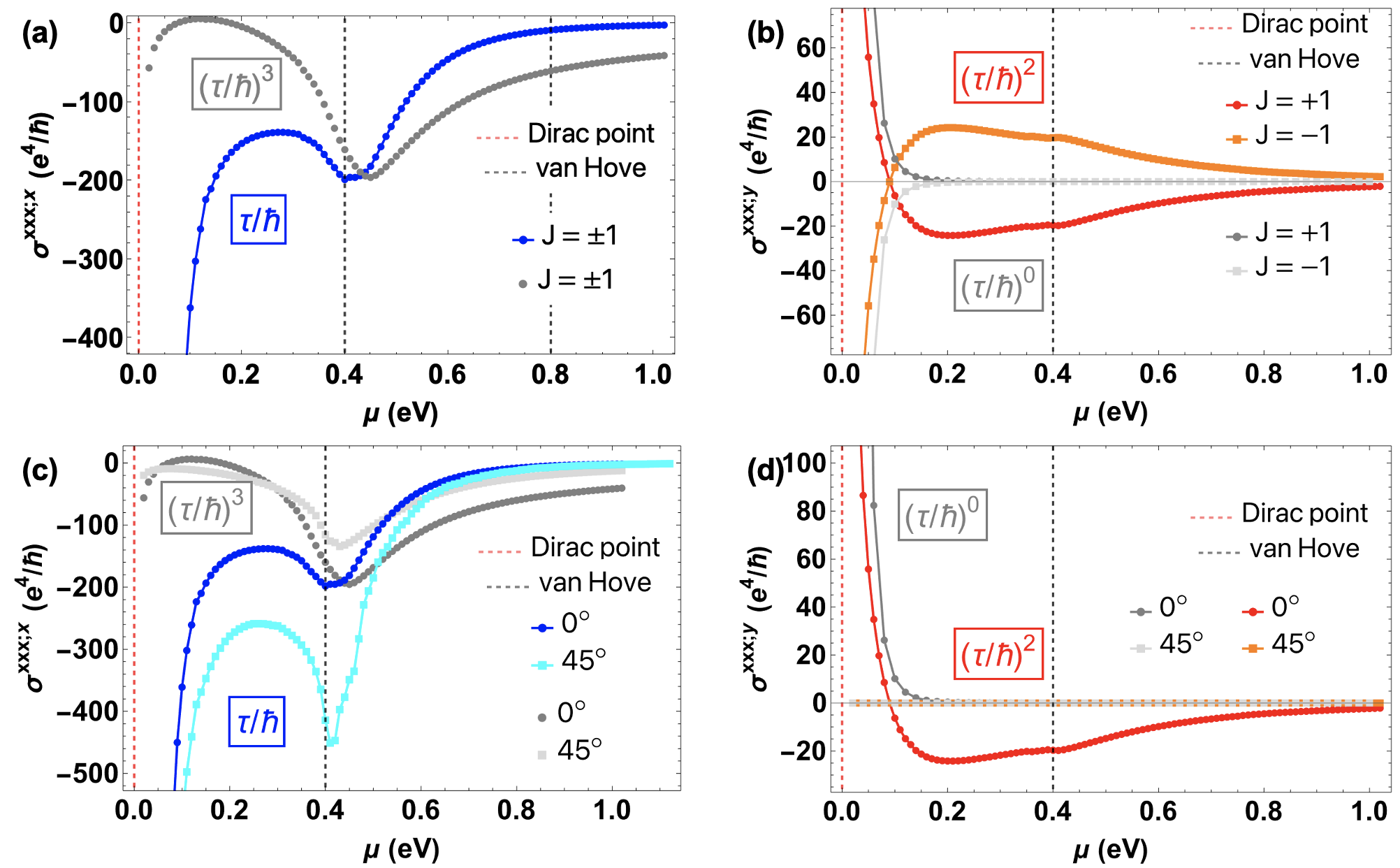}
    \caption{(a) Longitudinal conductivity is even under $J \rightarrow -J$.
    (b) Hall conductivity is odd under $J \rightarrow -J$. (c) Longitudinal conductivity under $45^\circ$ rotation. (d) Hall conductivity under $45^\circ$ rotation. We use parameters $J=\pm 1$eV, $\lambda=0.4$eV.}
    \label{fig:Fig3}
\end{figure}
\indent To experimentally verify that the response indeed originates from magnetic order, one may consider which components of the nonlinear conductivity are even/odd under $J \rightarrow -J$.
For both ferromagnets and AFMs, the sign of $J$
can be controlled by sweeping the external magnetic field \cite{jiang2018electric,QMDscience,QMDnature,gao2021layer}. 
We expect altermagnets may be manipulated similarly to other AFMs.
However, the anisotropic nature of altermagnets provides an even simpler way to switch the sign of the magnetic order, namely by a $90^\circ$ rotation. Unlike the second-order response in $\mathcal{PT}$-symmetric AFMs, where the QMD is $\cal T$-odd, here the QMQ is $\cal T$-even and instead the BCQ is $\cal T$-odd, as shown in Figs.~\ref{fig:Fig3}(a), (b). The AIC term is also $\cal T$-odd. Thus, the Hall response flips sign by switching the sign of $J$ (or by a $90^\circ$ crystal rotation) 
while the longitudinal part remains invariant, providing another experimentally verifiable prediction. 

The anistropic nature of $d$-wave altermagnets dictates that under $45^\circ$ rotation $d_{x^2-y^2}\rightarrow d_{xy}$. Figs.~\ref{fig:Fig3}(c), (d) show how  $\sigma^{xxx;x}$ and $\sigma^{xxx;y}$ vary under a $45^\circ$ rotation. While the longitudinal part remains finite with an enhanced QMQ, the Hall response vanishes
due to the Berry curvature of a $d_{xy}$-wave altermagnet being odd in $k^{x,y}$. 

Finally, we comment on two other components of the nonlinear conductivity:  $\sigma^{xxy;x}$ and $\sigma^{xyy;x}$. The first is proportional to the Hall response, as shown in \cite{sm}. For a $d_{x^2-y^2}$-wave altermagnet the $\tau^2$-dependent BCQ contribution to $\sigma^{xyy;x}$ vanishes and thus, similar to $\sigma^{xxx;x}$, $\sigma^{xyy;x}$ is only comprised of nonlinear Drude and QMQ terms. While measuring mixed conductivities is  challenging, they provide further benchmarks for future experiments, e.g., the sum frequency generation method \cite{QMDscience}. \\
\indent \blue{\emph{Third-order response in RuO$_2$}}.-- While the symmetry arguments apply to 3D planar altermagnets, the peaks in conductivity are most prominent for strictly 2D systems. 
Nevertheless, we predict that weaker signatures persist in a thin slab of RuO$_2$, as we now describe.

We consider a thin slab where the surface normal is parallel to the Neel vector $\mathcal{N}$, thus preserving $C_{4z}\mathcal{T}$. 
\begin{figure}[t]
    \centering
    \includegraphics[width=\linewidth]{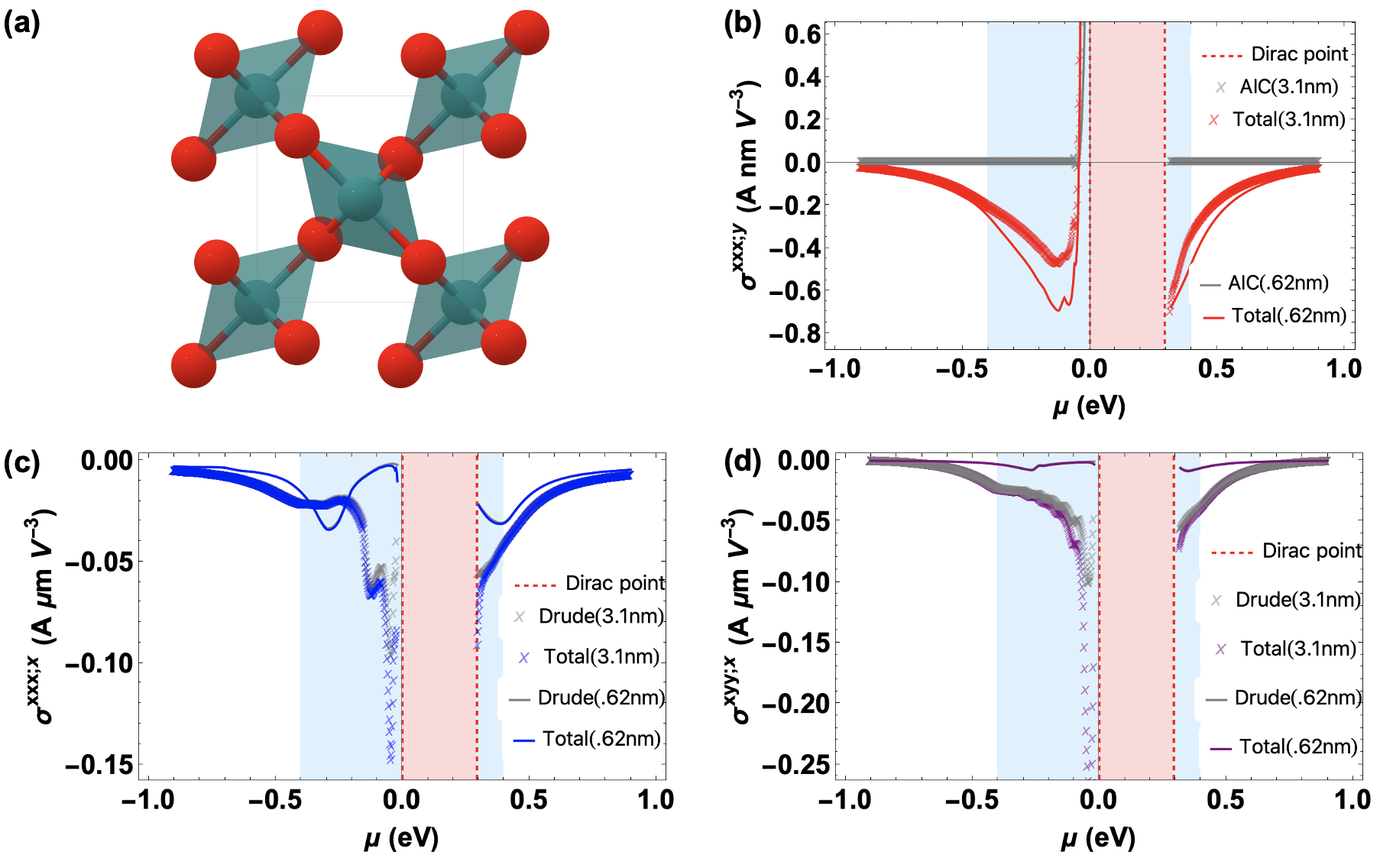}
    \caption{ 
    (a) Unit cell of RuO$_2$~\cite{osti_1307989}. Ru atoms (green) are related by four-fold screw symmetry. The O atoms (red) provide an anisotropic potential for the Ru atoms. 
    The nonlinear conductivities for scattering time $\hbar/\tau = 0.1$eV are shown in (b) $\sigma^{xxx;y}$ (c) $\sigma^{xxx;x}$ and (d) $\sigma^{xyy;x}$ for thickness $d=0.62$nm and $d=3.1$nm. For the details on the model and paramters see \cite{sm}.
    The blue shaded region indicates the anti-crossing contribution. The pink shaded region indicates the divergent region contributed by Dirac points.}
    \label{fig:Fig4}
\end{figure}
Figs.~\ref{fig:Fig4}(b), (c), show the nonlinear longitudinal and transverse responses for thin films of RuO$_2$ with thickness $d=0.62$ nm and $d=3.1$ nm (see \cite{sm} for details on the thin film model). For $\hbar/\tau=0.1$ eV~\cite{PhysRevLett.127.277201},
the AIC term is negligible and third-order Hall is dominated by BCQ, as shown in Fig.~\ref{fig:Fig4}(b). Also, as expected, the anti-crossing contribution is more pronounced for thinner samples, i.e., in the 2D limit. 
For an electric field of $1$ V/$\mu$m, a sample size of around $1 \text{mm}\times 1\text{mm} \times 3.1 \text{nm}$ and $\mu \sim 0.4 $eV, we find a current of $2$A and $1.5$mA for longitudinal and Hall responses, respectively.\\
\indent 
\indent \blue{\emph{Conclusion}}.-- Our symmetry analysis unveils the quantum geometric origin of the nonlinear response in altermagnets. Unlike $\mathcal{PT}$-symmetric antiferromagnets, the leading order response is third order and the QMQ plays an indispensable role.
Our results not only provide unique features of altermagnets in nonlinear transport, but also greatly expand the material platforms in which quantum geometric multipoles can be directly probed. \\
\indent \blue{\emph{Acknowledgements}}.---This work was supported by the Air Force Office of Scientific Research under Grant No. FA9550-20-1-0260. J.C. is partially supported by the Alfred P. Sloan Foundation through a Sloan Research Fellowship. The Flatiron Institute is a division of the Simons Foundation. Y.F. is partially supported by the U.S. DOE, BES, under Award No. DE-SC0018197, and the AFOSR under Grant No. FA9550-21-1-0356.

\bibliography{reference}

\newpage \clearpage

\onecolumngrid
\setcounter{secnumdepth}{3}
\appendix

\begin{center}
	{\large
Quantum geometry induced nonlinear transport in altermagnets
	\vspace{4pt}
	\\
	SUPPLEMENTAL MATERIAL
	}
\end{center}

\section{Derivation of nonlinear conductivities}
There are several methods to derive the nonlinear conductivities~\cite{PhysRevLett.112.166601,PhysRevB.106.035307,mckay2023spatially,parker2019diagrammatic,LawPRB2023,PhysRevB.107.075411,PhysRevB.107.205120,PhysRevB.105.045118,PhysRevLett.127.277201,sipe2000second,blount1962bloch,kaplan2023unifying,kaplan2022unification}, which mainly fall into two classes: (1) semi-classically defining current $\mathbf j=-e\mathbf v$ and solving the perturbation problem $H\mapsto H-e\mathbf E \cdot \mathbf r$; or (2) applying Peierls substitution $H(\mathbf k)\mapsto H(\mathbf k+e\mathbf A)$ and defining current as $\mathbf j=\frac{\delta H}{\delta \mathbf A}$. 
In this work, we will use approach (1) to determine the third order responses as it provides a clear connection between the nonlinear conductivity and the quantum geometric tensor multipoles.

\subsection{Nonlinear conductivity in the semi-classical approach}
The current of electrons in a lattice is given by
\begin{align}
    {\mathbf j} = -e \int_{k} \sum_{n} f_n {\mathbf v}_n,
\end{align}
where $n$ is denoting the $n$-th band, $f_n$ is the electron density of the $n$-th band (at equilibrium without external fields $f_n = \frac{1}{e^{\beta (\varepsilon_n-\mu)}+1}$, the Fermi-Dirac distribution), $\varepsilon_n$ is the energy of the $n$-th band, $\mu$ is the chemical potential and the group velocity is given by the traditional group velocity plus the anomalous velocity
\begin{equation}
    {\mathbf v}_n = \frac{1}{\hbar}\frac{\partial \varepsilon_n}{\partial {\mathbf k}} - \frac{e}{\hbar} {\mathbf E} \times {\mathbf \Omega_n},
\end{equation}
where $\mathbf\Omega_n$ is the Berry curvature of the $n$-th band and $\mathbf E$ is the external electric field. Note: the Berry curvature and velocity can only be separately defined for each band when the bands are all non-degenerate.

The $l$-th order conductivity responses are defined for current $j^{(l)}$ generated by $l$ powers of $\mathbf E$ as
\begin{equation}
\label{eqn:SM_sig_def}
     \sigma^{a_1\dots a_l;d} = \frac{1}{l!} \frac{j^{(l)}_d}{E^{a_1}\dots  E^{a_l} }= \frac{1}{l!} \frac{\partial^lj_d}{\partial E^{a_1}\dots \partial E^{a_l} }\bigg|_{\mathbf E=0}
\end{equation}
where $a_1,\dots ,a_l, d = x,y,z$ and $j$ is the total current (all orders). Expanding in powers of $\mathbf E$:
\begin{align}
    f_n &= f^{(0)}_n + f^{(1)}_n + f^{(2)}_n + f^{(3)}_n +\dots  \\
    {\mathbf v}_n &= {\mathbf v}^{(0)}_n + {\mathbf v}^{(1)}_n + {\mathbf v}^{(2)}_n + {\mathbf v}^{(3)}_n + \dots ,
\end{align}
which yields the $l$-th order electron current for $l=0,1,2,3$ 
\begin{align}
    \mathbf j ^{(0)} &= -e \int_k \sum_n f_n^{(0)} \mathbf v_n^{(0)} = 0 \\
    \mathbf j ^{(1)} &= -e \int_k \sum_n f_n^{(1)} \mathbf v_n^{(0)} + f_n^{(0)} \mathbf v_n^{(1)} \label{eqn:SMj1} \\
    \mathbf j ^{(2)} &= -e \int_k \sum_n f_n^{(2)} \mathbf v_n^{(0)} + f_n^{(1)} \mathbf v_n^{(1)} + f_n^{(0)} \mathbf v_n^{(2)} \label{eqn:SMj2}\\
    \mathbf j ^{(3)} &= -e \int_k \sum_n f_n^{(3)} \mathbf v_n^{(0)} + f_n^{(2)} \mathbf v_n^{(1)} + f_n^{(1)} \mathbf v_n^{(2)} + f_n^{(0)} \mathbf v_n^{(3)} \label{eqn:SMj3}
\end{align}
Thus, to determine the $l$-th order conductivity, we must compute $f$ and $v$ order by order in $\mathbf{E}$.

First let us determine the corrections to the Fermi-Dirac distributions. We consider the semi-classical Boltzmann equation of electrons under an external electric field $\mathbf E$
\begin{align}
    \partial_t f+\frac{\mathbf{F}}{\hbar} \cdot \nabla_{\mathbf{k}} f+\mathbf{v} \nabla_{\mathbf{r}} f&=\mathcal{I}(f),
    \end{align}
which yields
\begin{align}
    i\omega f+\frac{-e\mathbf{E}}{\hbar}  \cdot \nabla_{\mathbf{k}} f &=-\frac{f-f_0}{\tau}
\end{align}
after applying the relaxation time approximation to the collision rate and assuming $f$ is spatially uniform. Then $f_n^{(l)}$ can be obtained recursively
\begin{equation}
\label{eqn:SMfn}
    f_n^{(l)} = \frac{ \frac{e}{\hbar}{\mathbf E}^l \nabla_{\mathbf{k}} f_n^{(l-1)}}{i\omega + 1/\tau}
    = \left(\frac{ e/{\hbar}}{i\omega + 1/\tau} \right)^l{\mathbf E}^l \nabla_{\mathbf{k}}^l f_n^{(0)}
\end{equation}
where ${\mathbf E}^l \nabla_{\mathbf{k}} = E^{a_1}E^{a_2}\dots E^{a_l} \partial_{k^{a_1}}\partial_{k^{a_2}}\dots \partial_{k^{a_l}}$.
In the direct current limit, where $\omega \ll 1/\tau$, 
\begin{equation}
    \lim_{\omega \rightarrow 0} f_n^{(l)} \approx \left( \frac{e\tau}{\hbar} \right)^l{\mathbf E}^l \nabla_{\mathbf{k}}^l f_n^{(0)}
\end{equation}

Next we determine the expansion of $\mathbf v_n$. The group velocity has contributions from the band dispersion and the Berry curvature, as follows:
\begin{align}
    \mathbf v_n^{(0)} &=\frac{1}{\hbar} \frac{\partial \varepsilon_n^{(0)}}{ \partial {\mathbf k}} \\
    \mathbf v_n^{(1)} &= \frac{1}{\hbar}\frac{\partial \varepsilon_n^{(1)}}{ \partial {\mathbf k}} - \frac{e}{\hbar} {\mathbf E} \times {\mathbf \Omega_n^{(0)}} \\
    \mathbf v_n^{(2)} &= \frac{1}{\hbar}\frac{\partial \varepsilon_n^{(2)}}{ \partial {\mathbf k}} - \frac{e}{\hbar} {\mathbf E} \times {\mathbf \Omega_n^{(1)}} \\
    \mathbf v_n^{(3)} &=\frac{1}{\hbar} \frac{\partial \varepsilon_n^{(3)}}{ \partial {\mathbf k}} - \frac{e}{\hbar} {\mathbf E} \times {\mathbf \Omega_n^{(2)}}
\end{align}
where it is natural to define the expansions of $\epsilon$ and $\Omega$ in powers of $\mathbf E$
\begin{align}
    \varepsilon_n &= \varepsilon^{(0)}_n + \varepsilon^{(1)}_n + \varepsilon^{(2)}_n + \varepsilon^{(3)}_n +\dots\\
    \mathbf\Omega_n &= \mathbf\Omega^{(0)}_n + \mathbf\Omega^{(1)}_n + \mathbf\Omega^{(2)}_n +\dots
\end{align}
Evaluating these terms yields every order of the nonlinear conductivity.

\subsection{Perturbative expansions of energy and Berry curvature}
We apply a Schrieffer-Wolff transformation to perturbatively expand $\varepsilon_n^{(l)}$ and $\mathbf \Omega_n^{(l)}$ in powers of $\mathbf{E}$.
We write the Hamilitonian as
\begin{equation}
    H = \sum_{mn} \left( \varepsilon^{(0)}_n \delta_{nm} - e{\mathbf E}\cdot \langle n |{\mathbf r} | m\rangle  \right) | n\rangle \langle m |,
\end{equation}
where $|n\rangle$ is the unperturbed wavefunction of the $n$-th band eigenstate at zero field. For simplicity we assume there are no band degeneracies.
The Schrieffer-Wolff transformation requires the perturbation term to be off-diagonal. Therefore, we define 
\begin{align}
    H_0&=\sum_n \left(\varepsilon^{(0)}_n-eE^aA^a_n\right)|n\rangle\langle n| \\
    H_1 &= \sum_{n\neq m}\left(-eE^aA^a_{nm} \right)|n\rangle\langle m|
\end{align}
where $A_{n}=\langle n|{\mathbf r}|n\rangle$ and $A_{nm}=\langle n|{\mathbf r}|m\rangle$.
The Schrieffer-Wolff transformation prescribes that the perturbation of an operator $\cal O$ in powers of $S$ is given by
\begin{align}
    {\cal O} \longrightarrow e^S {\cal O} e^{-S} = {\cal O} + [{\cal O},S] + \frac12 [S,[S,{\cal O}]] + \frac16 [S,[S,[S,{\cal O}]]] + \dots
\end{align}
The operator $S$ is usually chosen so that the first order perturbation of $H_0$ cancels the zero-th order of $H_1$, diagonalizing the Hamiltonian  to first order, i.e.
\begin{equation}
    H_1+[S,H_0] = 0
\end{equation}
In our case, we solve for $S$ as 
\begin{equation}
    S_{nn}=0, \quad S_{nm} = \frac{-eE^a A^a_{nm}}{\varepsilon_{nm}-e{\mathbf E}\cdot ({\mathbf A}_n-{\mathbf A}_m)} \approx \frac{-eE^a A^a_{nm}}{\varepsilon_{nm}}- \frac{e^2E^a E^b A^a_{nm}(A^b_{n}-A^b_{m})}{\varepsilon_{nm}^2}
\end{equation}
where $\varepsilon_{nm} = \varepsilon_n^{(0)} -\varepsilon_m^{(0)}$ and 
the second term is a higher-order correction that we will only use in calculating $\mathbf v^{(3)}_n$. For the other terms we only need to keep the first term $S_{nm} \approx {-eE^a A^a_{nm}}/{\varepsilon_{nm}}$~\cite{kaplan2022unification}.

Then the Hamiltonian in the new basis is
\begin{align}
    H\longrightarrow H' &= H_0+(H_1+[S,H_0])+([S,H_1]+\frac12[S,[S,H_0]]) + (\frac12[S,[S,H_1]]+\frac16[S,[S,[S,H_0]]]) \dots \nonumber\\
    &=H_0+([S,H_1]+\frac12[S,[S,H_0]])+ (\frac12[S,[S,H_1]]+\frac16[S,[S,[S,H_0]]]) +\dots \nonumber \\
    &=H_0+\frac12[S,H_1] + \frac13 [S,[S,H_1]] +\dots
\end{align}
The perturbed $\varepsilon_n$ is given by $\varepsilon_n = \langle n| H' |n\rangle$. Expanding $\varepsilon_n$ in orders of $\mathbf E$ yields the corrections
\begin{align}
    \varepsilon_n^{(1)} &= -eE^a A_n^a\\
    \varepsilon_n^{(2)} &=  \frac12e^2E_aE_b \left( \sum_{m\neq n} \frac{A_{nm}^a A_{mn}^b+A_{mn}^a A_{nm}^b}{\varepsilon_{nm}}\right) \equiv e^2G_n^{ab}E_aE_b \label{eqn:SME2}\\
    \varepsilon_n^{(3)} &= - e^3 E_aE_bE_c \left( \sum_{m\neq n} \sum_{l\neq m, n} \frac{A_{nl}^a A_{lm}^bA_{mn}^c }{\varepsilon_{nm}\varepsilon_{nl}}\right) + e^3E_aE_bE_c \left(\sum_{m\neq n} \frac{A_{nm}^aA_{mn}^b(A^c_n-A^c_m)}{\varepsilon_{nm}^2}\right) \\
    &\xrightarrow[]{\text{two band limit}} e^3E_aE_bE_c \left( \frac{A_{n\bar n}^aA_{\bar n n}^b(A^c_n-A^c_{\bar n})}{\varepsilon_{n\bar n}^2}\right) \\
    & = e^3E_aE_bE_c \left( G_n^{ab} \frac{A^c_n-A^c_{\bar n}}{\varepsilon_{n\bar n}}\right) \label{eqn:SME3}
\end{align}
where in the two band limit, $\bar{n}$ indicates the band that is not $n$.
The two band limit is a good approximation when there is only one other band $\bar n$ that has a small energy gap with respect to the band $n$. For the all the other bands that have large gaps their contribution is negligible. 
The term $\varepsilon_n^{(1)}=\mathbf E \cdot \mathbf P_n$, $\mathbf P_n=-e \mathbf A_n=-e\langle n|\mathbf r|n\rangle$ is the electron polarization of the $n$-th band at a momentum $\mathbf k$; the term $\varepsilon_n^{(2)}$ is corrected by the band-normalized quantum metric $G_n^{ab}=\frac{A_{nm}^a A_{mn}^b+A_{mn}^a A_{nm}^b}{2\varepsilon_{nm}}$; and the term $\varepsilon_n^{(3)}$ will contribute to the AIC term discussed in the main text.

The Berry connection $\mathbf A_n(\mathbf k) = \langle n|\mathbf r|n\rangle$ is gauge dependent. Thus, certain terms, such as $\varepsilon^{(1)}_n$, are not well-defined. This occurs because the Hamiltonian $H^{ext} = -e\mathbf E \cdot \mathbf r$ breaks translation symmetry and the Bloch states are no longer well defined. 
To continue with the semi-classical approach, we should drop the unphysical gauge dependent terms by choosing a gauge that sets $\mathbf E \cdot \mathbf A_n=0$.
Then for this choice of gauge and in the two band limit $\varepsilon_n^{(1)}=\varepsilon_n^{(3)}=0$,
and thus the AIC contribution from $\varepsilon_n^{(3)}$ is negligible. 

Next, We expand the Berry curvature in powers of $\mathbf{E}$ by applying the Schrieffer-Wolff transformation to the Berry connection $A_n$:
\begin{align}
    \mathbf A \longrightarrow \mathbf A' &= \mathbf A+[S, \mathbf A]+\frac12[S,[S,\mathbf A]]\dots
\end{align}
and $\mathbf A_n=\langle n| \mathbf A' |n\rangle$. We expand $A'$ in powers of $\mathbf E$
\begin{align}
    \left(\mathbf A^{(1)}\right)^b &= -e E^a G_{n}^{ab}\\
    \left(\mathbf A^{(2)}\right)^c &= e^2E^aE^b  \left(\sum_{m\neq l,n} \sum_{l\neq m,n} \frac{A_{nl}^aA_{lm}^bA_{mn}^c}{\varepsilon_{nl}\varepsilon_{mn}}\right) - e^2E_aE_b \left(\sum_{m\neq n} \frac{A_{nm}^aA_{mn}^b(A^c_n-A^c_m)}{\varepsilon_{nm}^2}\right) \\
    &\xrightarrow[]{\text{two band limit}} - e^2E_aE_b \left( \frac{A_{n\bar n}^aA_{\bar n n}^b(A^c_n-A^c_{\bar n})}{\varepsilon_{n \bar n}^2}\right) \\
    &= - e^2E_aE_b \left( G^{ab}_n \frac{A^c_n-A^c_{\bar n}}{\varepsilon_{n \bar n}}\right) 
\end{align}
where in the two band limit, $\bar{n}$ indicates the band that is not $n$.
Then the Berry curvature correction is
\begin{align}
    \left( {\mathbf \Omega}_n^{(1)} \right)^c &= -e E_d \epsilon^{abc} \partial_a G^{bd}_n  \label{eqn:SMOmega1}\\
    \left( {\mathbf \Omega}_n^{(2)} \right)^c &= - e^2E_aE_b \epsilon^{cde} \partial_d \left( G^{ab}_n \frac{A^e_n-A^e_{\bar n}}{\varepsilon_{n \bar n}}\right) = - e^2E_aE_b \left( G^{ab}_n \frac{\Omega^c_n-\Omega^c_{\bar n}}{\varepsilon_{n \bar n}}\right) \label{eqn:SMOmega2}
\end{align}
where ${\mathbf \Omega}_n^{(2)}$ contributes to the AIC term. In the two-band model where two bands do not hybridize We can further simplify this term by noting, $\mathbf \Omega^c_{\bar n}=-\mathbf \Omega^c_{n}$. In the main text we show that for altermagnets, this term only contributes to the conductivity near the Dirac point and is insensitive to the anti-crossings near van Hove point.
 
Therefore, by taking in to account all the corrections to $\Omega$ up to second-order in $\mathbf{E}$, we have included all the relevant terms up to third-order in $\mathbf{E}$ to the conductivity in this approach.

\subsection{The nonlinear conductivity}
Combining Eqs.~\eqref{eqn:SMj1},\eqref{eqn:SMj2},\eqref{eqn:SMj3},\eqref{eqn:SMfn},\eqref{eqn:SME2},\eqref{eqn:SME3},\eqref{eqn:SMOmega1}, and \eqref{eqn:SMOmega2}, we are ready to write down the nonlinear conductivity, following the definition  Eq.~(\ref{eqn:SM_sig_def}).
Note that the permutation symmetry of electric fields are automatically satisfied.
For example, when calculating the second order conductivity, we get a term $j^c \propto E^aE^b \partial_a f~\Omega^{bc}$. The conductivity from this term is $\sigma^{ab;c}= \frac12\frac{\partial^2 j^c}{\partial E^a \partial E^b}\big|_{\mathbf E=0}\propto \frac12(\partial_af~\Omega^{bc}+\partial_bf~\Omega^{ac})$, where $a$ and $b$ are symmetric and there is a coefficient $1/2$.

After symmetrization, we arrive at the final expressions for the nonlinear conductivity
\begin{align}
    \sigma^{a;b} &= \frac{e^2}{\hbar} \int_k \sum_n f_n \Omega^{ab}_n - \frac{e^2\tau}{\hbar^2} \int_k \sum_n \frac{\partial f_n}{\partial k_a}\frac{\partial\varepsilon_n}{\partial k_b} \\
    &=\frac{e^2}{\hbar} \sum_n \int_k  f_n \Omega^{ab}_n + \frac{e^2\tau}{\hbar^2} \sum_n \int_k f_n \frac{\partial^2\varepsilon_n}{\partial k_a \partial k_b} 
\end{align}

\begin{align}
    \sigma^{a b ; c}= 
& -\frac{e^3 \tau^2}{\hbar^3} \sum_n \int_{\boldsymbol{k}} f_n \partial_{k^a} \partial_{k^b} \partial_{k^c} \varepsilon_n \\
& +\frac{e^3 \tau}{\hbar^2} \sum_n \int_{\boldsymbol{k}} f_n \frac12 \left(\partial_{k^a} \Omega_n^{b c}+\partial_{k^b} \Omega_n^{a c}\right) \\
& -\frac{e^3}{\hbar} \sum_n \int_{\boldsymbol{k}} f_n\left(2 \partial_{k^c} G_n^{a b}-\frac{1}{2}\left(\partial_{k^a} G_n^{b c}+\partial_{k^b} G_n^{a c}\right)\right)
\end{align}

\begin{align}
\label{eqn:SM3rd}
    \sigma^{abc;d} =
    & \frac{e^4 \tau^3}{\hbar^4} \sum_n \int_{\boldsymbol{k}} f_n \partial_{k^a} \partial_{k^b} \partial_{k^c} \partial_{k^d} \varepsilon_n \\
    & -\frac{e^4 \tau^2}{\hbar^3} \sum_n \int_{\boldsymbol{k}} f_n \frac13 \left(\partial_{k^a} \partial_{k^b} \Omega_n^{c d}+\partial_{k^b} \partial_{k^c} \Omega_n^{a d}+\partial_{k^a} \partial_{k^c} \Omega_n^{b d}\right) \\
    & +\frac{e^4\tau}{\hbar^2} \sum_n \int_{\boldsymbol{k}} f_n \frac13\Big(2 ( \partial_{k^a} \partial_{k^d} G_n^{bc}+\partial_{k^b} \partial_{k^d} G_n^{ac}+\partial_{k^c} \partial_{k^d} G_n^{ab} ) \nonumber \\
    &\hspace{3cm} -\left(\partial_{k^a}\partial_{k^c} G_n^{b d}+\partial_{k^b}\partial_{k^c} G_n^{a d}+\partial_{k^a}\partial_{k^b} G_n^{cd} \right) \Big) \\
    &+\text{AIC},
\end{align}
where the additional interband contribution AIC is
\begin{equation}
    \text{AIC} = -\left[\frac{e^4}{\hbar}\sum_n \int_{\boldsymbol k}f_n \left(\frac{2}{3\varepsilon_{n\bar n}} \left( G^{ab}_n \Omega^{cd}_n + G^{ac}_n \Omega^{bd}_n + G^{bc}_n \Omega^{ad}_n \right) \right) \right]
\end{equation}

\section{Symmetry classification of Berry curvature and quantum metric multipoles}
In this section, we first give a brief review of spin groups. Then we discuss the classification of Berry curvature multipoles and quantum metric multipoles for general groups with both the real space and spin space elements.

\subsection{Spin groups}
Spin groups describe the symmetries of certain magnetically ordered materials without spin-orbit coupling (SOC), which can be described by the following Hamiltonian~\cite{litvin1974spin,litvin1977spin}
\begin{equation}
    \label{eqn:HamSG}
    H = H_0 + \mathbf M(\mathbf r)\cdot \mathbf{\sigma}
\end{equation}
where $H_0$ is the Hamiltonian describing the lattice and orbital degrees of freedom and the Zeeman like term describes the interaction between orbitals and spins. Here $\mathbf M(\mathbf r)$ is a pseudo-vector that determines the spin orientation and $\sigma=(\sigma_x,\sigma_y,\sigma_z)$ is the Pauli matrix vector. Eq.~(\ref{eqn:HamSG}) is preserved by a spin group, where each symmetry operation is a combination of a real space symmetry and a spin space symmetry, taking the form $[g_s||g]$ 
where $g_s$ acts on the spin space while $g$ acts on the real space. The spin parts $g_s$ form a group denoted by $S$, while the lattice parts $g$ forms a group denoted by $L$. 
The Hamiltonian $H_0$ in Eq.~(\ref{eqn:HamSG}) is invariant under the space group $L$ in addition to the $SU(2)$ spin symmetry and time-reversal symmetry; thus, it is invariant under  $L\times SU(2)\times \mathbb Z_2^T$.

For every spin group there must exist a normal subgroup $L_0$ of $L$ and a normal subgroup $S_0$ of $S$ such that~\cite{litvin1974spin} 
\begin{equation}
    \frac{L}{L_0} \cong \frac{S}{S_0}.
\end{equation}
The relation between spin groups and magnetic groups can be understood from the perspective of symmetry breaking~\cite{PhysRevX.12.021016}.
Adding Zeeman-like terms to $H_0$ breaks the symmetry from the space group $L$ to the spin group $G$. SOC will break it down further, resulting in a magnetic space group $L'$. To summarize: the original Hamiltonian $H_0$ is invariant under symmetries of the form $[E_s||g] \in L$, where $E_s$ is the identity in spin space and $g\in L $, in addition to rotations in spin space in $SU(2)$ and time-reversal symmetry $\cal T$; the generalized Zeeman term is invariant under symmetries of the form $[g_s||g] \in G$; and a possible SOC term, not included in Eq.~\eqref{eqn:HamSG}, is invariant under symmetries of the form $[g||g], [g||g]{\cal T} \in L'$. 
We summarize this view as
\begin{equation}
    L\times SU(2)\times {\mathbb Z}_2^T \xrightarrow{ \mathbf M(\mathbf r)\cdot \mathbf \sigma} G  \xrightarrow{\text{SOC}} L'
\end{equation}
Note that both $G$ and $L'$ describe the symmetries of magnetically ordered systems, while $L\times SU(2)$ does not.

Given a spin group element $[g_s || g ]$,
the spin space symmetry $g_s$ only acts on the pseudo-vector $\mathbf M$ in Eq.~(\ref{eqn:HamSG}). 
The spin group element $\hat g = [g_s||g]$ acts on the Hamiltonian as,
\begin{align}
    \hat g : H_0 &\mapsto g H_0 g^{-1}\\
    \hat g : \mathbf M(\mathbf r)\cdot \sigma &\mapsto \left(s(g_s) \mathbf M(R_{g}{}^{-1}\mathbf r) \right) \cdot \sigma 
    \label{eqn:transM}
\end{align}
where $R_g$ is a vector representation and $s(.)$ is a pseudo-vector representation, since $\mathbf r$ is a vector while $\mathbf M$ is a pseudo-vector. There are two equivalent ways that a spin group element acts on the altermagnetic order: (i) via rotating the the spatial part of the altermagnetic order $\mathbf{M}$ as shown in \eqref{eqn:transM}, or (ii) through rotation of the spin part $\sigma$ as, 
\begin{equation}
\label{eqn:rep_gs}
    \hat{g}: \mathbf M(\mathbf r) \cdot \mathbf \sigma \mapsto \mathbf M(R_g^{-1}\mathbf r) \cdot \tilde{s}(g_s^{}) \mathbf \sigma  \tilde{s}(g_s^{-1}) =: s(g_s)\mathbf M (R_g^{-1}\mathbf r)\cdot \mathbf \sigma
\end{equation}
where $\tilde s$ is a spin-$1/2$ representation of $SU(2)$ while $s$ is a pseudo-vector representation of $SO(3)$. Since $\tilde{s}(g_s)$ is a natural representation of a spin space symmetry, the representation $s(g_s)$ should be understood as a definition. 
The latter choice, Eq.~\eqref{eqn:rep_gs}, is more intuitive because it shows directly how the spin group acts on the quantum geometric tensor.

\subsection{Classification of Berry curvature and quantum metric multipoles for spin space groups}

The quantum metric and Berry curvature are the real symmetric part and imaginary anti-symmetric part of the quantum geometric tensor defined by ${\mathcal{G}}_{a b} = \tr(P\partial_a P\partial_b P) = \mathfrak{g}_{a b} - \frac i2 \Omega_{a b}$ where $P$ is the projector to the bands of interest.
In general, the quantum metric $\mathfrak{g}_{a b} $ transforms as a rank-2 symmetric tensor under spatial symmetries while Berry curvature transforms as a rank-2 antisymmetric tensor which is dual to a pseudo-vector $\Omega^{c}\equiv\epsilon^{abc} \Omega_{a b} $ in three dimensional space.

First, we need to know how the derivatives $\partial_{\boldsymbol{k}}$ transform under symmetries, since they appear in the quantum metric multipoles, Berry curvature multipoles and Drude terms. Each derivative transforms under the symmetry $[g_s||g]$ as
\begin{equation}
\label{eqn:dkunderg}
\partial_{{\left(R_{g}\boldsymbol{k}\right)}^{c'}} = R_{g}{}^{c'}_{c} \partial_{\boldsymbol{k}^{c}}
\end{equation}

Then we consider how quantum geometric tensor transforms under symmetries.
Under a general unitary symmetry operator $\hat{g}=[g_s||g]$, the projector transforms as $P(R_g{\mathbf k})=\hat{g} P({\mathbf k})\hat{g}^\dagger$. We now show only the real space part $g$ transforms the quantum geometric tensor because the derivatives in the definition of $\mathcal{G}$ are purely spatial
\begin{align}
    {\mathcal{G}}_{a b}(R_g{\mathbf k}) &= \tr\left( P(R_g{\mathbf k}) \partial_a P(R_g{\mathbf k})\partial_b P(R_g{\mathbf k}) \right) \cr
    &= \tr\left( \hat{g}P({\mathbf k})\hat{g}^\dagger \left( \partial_a  \hat{g}P({\mathbf k}) \hat{g}^\dagger \right) \left( \partial_b \hat{g}P({\mathbf k})\hat{g}^\dagger \right) \right) \cr
    &= \tr\left( P({\mathbf k}) (\hat{g}^\dagger \partial_a \hat{g}) P({\mathbf k}) (\hat{g}^\dagger \partial_b \hat{g}) P({\mathbf k}) \right) \cr
    &={\mathcal{G}}_{(R_g{\mathbf k})^a,(R_g{\mathbf k})^b}(\mathbf k) \cr
    &=R_{g}{}_{a}^{a'}R_{g}{}_{b}^{b'} {\mathcal{G}}_{a' b'}(\mathbf k)
\end{align} 
The real space symmetry part $g=\{R_g|\boldsymbol{t}_g\}$ may contain a translation $\boldsymbol{t}_g$. The translational part is represented by $e^{i{\boldsymbol k}\cdot \boldsymbol{t}_g}$ in momentum space. It does not transform the quantum geometric tensor because it cancels out, i.e., $\partial_a e^{i{\boldsymbol k}\cdot \boldsymbol{t}_g} P e^{-i{\boldsymbol k}\cdot \boldsymbol{t}_g} = \partial_a P$.

For an anti-unitary symmetry $\hat{g} {\cal K}$, where $\hat g$ is unitary and $\cal K$ is the complex conjugation operator, the projection $P$ transforms as $P(R_g{\mathbf k})=\hat{g}{\cal K} P({\mathbf k}){\cal K}\hat{g}^\dagger = \hat{g} P({\mathbf k})^*\hat{g}^\dagger $. Thus the quantum geometric tensor transforms as 
\begin{align}
    {\mathcal{G}}_{a b}(R_g{\mathbf k}) &= \tr\big( P(R_g{\mathbf k}) \partial_a P(R_g{\mathbf k})\partial_b P(R_g{\mathbf k}) \big) \cr
    &= \tr\big( \hat{g}P({\mathbf k})^*\hat{g}^\dagger \partial_a  \hat{g}P({\mathbf k})^* \hat{g}^\dagger \partial_b \hat{g}P({\mathbf k})^*\hat{g}^\dagger \big) \cr
    &=R_{g}{}_{a}^{a'}R_{g}{}_{b}^{b'} {\mathcal{G}}^*_{a' b'}(\mathbf k)
\end{align} 

Since the quantum metric $\mathfrak g_{a b}(\boldsymbol{k})$ and Berry curvature $\Omega^{c'}(\boldsymbol{k})$ are the real symmetric and imaginary antisymmetric parts of the quantum geometric tensor, they transform as
\begin{align}
    \mathfrak g_{a' b'}(R_{g}{\mathbf k}) &= R_{g}{}_{a'}^{a}R_{g}{}_{b'}^{b} {\mathfrak g}_{a b}(\boldsymbol{k}) \\
    \Omega_{a'b'}(R_{g}\boldsymbol{k}) &= (-)^{\chi(\hat{g})}  R_{g}{}_{a'}^{a}R_{g}{}_{b'}^{b} \Omega_{ab}(\boldsymbol{k}) 
\end{align}
where $\chi(\hat{g})=\pm 1$ for a unitary/antiunitary symmetry $\hat{g}$.
Also, the dual Berry curvature $\Omega^c=\epsilon^{abc}\Omega_{ab}$ transforms as
\begin{align}
    \Omega^{c'}(R_{g}\boldsymbol{k}) &= \epsilon^{a'b'c'} \Omega_{a'b'}(R_g\boldsymbol{k})  \cr
    &= \det(R_g) R_{g}{}_{a}^{a'}R_{g}{}_{b}^{b'}R_{g}{}_{c}^{c'} \epsilon^{abc} ~(-)^{\chi(\hat{g})} R_{g}{}_{a'}^{d}R_{g}{}_{b'}^{e} \Omega_{de}(\boldsymbol{k})  \cr
    &= \det(R_g)R_{g}{}_{c}^{c'} \epsilon^{abc} ~(-)^{\chi(\hat{g})} \delta^d_a \delta^e_b \Omega_{de}(\boldsymbol{k})  \cr
     &= (-)^{\chi(\hat{g})} \det(R_g) ~ R_{g}{}^{c'}_{c} \Omega^{c}(\boldsymbol{k})
\end{align}
where $\det(R_g)=\pm 1$ for proper and improper rotations. 
Since the Levi Civita symbol $\epsilon^{abc}$ is a pseudovector, the term $\det(R_g)$ is required to satisfy the right-hand rule. It follows that the dual Berry curvature $\Omega^c$ is a also pseudo-vector.
From the above transformation, time-reversal symmetry maps $ \Omega^{c}(\boldsymbol{k}) \xrightarrow{\cal T}  -\Omega^{c}(-\boldsymbol{k}) $; inversion symmetry maps $ \Omega^{c}(\boldsymbol{k}) \xrightarrow{\cal I}   \Omega^{c}(-\boldsymbol{k}) $; and both time-reversal and inversion map $\mathfrak g_{ab}(\mathbf k)\mapsto \mathfrak g_{ab}(-\mathbf k)$.

Symmetry constrains the Fermi surface integrals of quantum geometric tensors and their multipoles.  
A symmetry $g$ forces the quantity $\int_k \sum_n f_n Q^n_{\alpha_1,\dots,\alpha_n,\gamma}$ to vanish if and only if
\begin{equation}
\label{eqn:multipole_condition}
     Q^n_{\alpha_1,\dots,\alpha_n,\gamma}(\boldsymbol{k}) \xrightarrow{g} - Q^n_{\alpha_1,\dots,\alpha_n,\gamma}(R_{g}\boldsymbol{k})
\end{equation}
since $f_n$ transforms trivially under symmetry. Now, following the above mentioned symmetry arguments we determine non-vanishing components of third order conductivity  for planar altermagnets.

In Table~\ref{tab:PlanarAlt} we list several symmetry groups that describe planar altermagnets~\cite{altermagnet1}, along with their corresponding magnetic point group.
Below we indicate in which magnetic groups the longitudinal and transverse conductivities survive/vanish.
The last column shows the sources of the transverse conductivity. We find for all the planar altermagnets that the longitudinal conductivity 
only has Drude and QMQ contributions. The longitudinal and transverse conductivities are given below where the BCQ and QMQ contributions to the transverse conductivity are separated:
\begin{align}
    \sigma^{xxx;x} & = \tau^3 \frac{e^4}{\hbar^4}\int \sum_nf_n \partial_x\partial_x\partial_x\partial_x\varepsilon_n + \tau \frac{e^4}{\hbar^2}\int \sum_nf_n  \partial_x\partial_x G^{xx}_n \label{eqn:SM_sigxxxx} \\
    \sigma^{xxx;y}_{BCQ} & = -\tau^2 \frac{e^4}{\hbar^3}\int \sum_nf_n \partial_x\partial_x\Omega^{xy}_n - \frac{e^4}{\hbar}\int \sum_nf_n \frac{2G^{xx}_n\Omega^{xy}_n}{\varepsilon_{n\bar n}} \label{eqn:SM_sigxxxyBCQ} \\
    \sigma^{xxx;y}_{QMQ} & = -\tau^2 \frac{e^4}{\hbar^3}\int \sum_nf_n \left( 2\partial_x\partial_yG^{xx}_n-\partial_x\partial_xG^{xy}_n \right) \label{eqn:SM_sigxxxyQMQ}
\end{align}
where the exact expression of the AIC term (the second term) in Eq.~(\ref{eqn:SM_sigxxxyBCQ}) is only valid for the two band limit and $\bar{n}$ denotes that band not indicated by $n$.
Table~\ref{tab:PlanarAlt} is obtained based on the above equations and the symmetry classification of BCQ and QMQ in the magnetic groups shown in Tables~\ref{tab:constraint2}, \ref{tab:constraint4}, \ref{tab:constraint41}, \ref{tab:constraint42}, \ref{tab:constraint43} and \ref{tab:constraint6}. 
As an example, $d$-wave altermagnets such as RuO$_2$, MnF$_2$ and MnO$_2$ belong to spin point group ${}^24/{}^1m{}^1m{}^2m$ and the magnetic point group $4'/mm'm$ in the presence of SOC as is classified by Table~\ref{tab:constraint41}.
\begin{table}[H]
    \centering
    \begin{tabular}{c|c|c|c|cc}
    \toprule
         Spin-momentum locking &Spin group & Magnetic group & Longitudinal & Transverse &Source
         \\
    \toprule
         & ${}^2m{}^2m{}^1m$ & $mmm$ & \cmark & \xmark & \\
         & ${}^24/{}^1m$ & $4'/m$ & \cmark & \cmark & BCQ+QMQ\\
    $d$-wave & ${}^24/{}^1m{}^1m{}^2m$  & $4'/mm'm$ & \cmark  & \cmark &BCQ\\
         & ${}^24/{}^1m{}^2m{}^1m$ & $4'/mmm'$ & \cmark & \xmark & \\
    \hline
    $g$-wave & ${}^14/{}^1m{}^2m{}^2m$ & $4/mmm$ & \cmark & \xmark &  \\
    \hline
    $i$-wave & ${}^16/{}^1m{}^2m{}^2m$ & $6/mmm$ & \cmark & \cmark &QMQ \\
    \botrule
    \end{tabular}
    \caption{Planar altermagnets, their symmetry groups, and the third order nonlinear longitudinal/transverse conductivity. 
    A (check)cross indicates that the conductivity is (non-)vanishing.
    When the transverse conductivity is non-vanishing the source of this conductivity is listed in the last column.} \label{tab:PlanarAlt}
\end{table}
\begin{table}[H]
    \centering
    \begin{tabular}{|c||c|c|c|c|}
    \hline
         $\int_{\mathbf k} \sum_n f_n$ &$\mathfrak{g}_{11}$ &$\mathfrak{g}_{22}$ &$\mathfrak{g}_{12}$ &$\Omega_{12}$ \\
    \hline
    \hline
         &$\mathfrak{g}_{11}$ &$\mathfrak{g}_{22}$ &$0$ &$0$ \\
    \hline
         $\partial_1$ &$0$ &$0$ &$0$ &$0$ \\
    \hline
         $\partial_2$ &$0$ &$0$ &$0$ &$0$ \\
    \hline
        $\partial_1\partial_1$ &$\partial_1\partial_1\mathfrak{g}_{11}$ &$\partial_1\partial_1\mathfrak{g}_{22}$ &$0$ &$0$ \\
    \hline
        $\partial_2\partial_2$ &$\partial_2\partial_2\mathfrak{g}_{11}$ &$\partial_2\partial_2\mathfrak{g}_{22}$ &$0$ &$0$\\
   \hline
        $\partial_1\partial_2$ &$0$ &$0$ &$\partial_1\partial_2 \mathfrak{g}_{12}$ &$\partial_1\partial_2 \Omega_{12}$ \\
    \hline
    \end{tabular}
    \caption{Constraint on quantum metric $\mathfrak{g}_{ij}$ multipoles and Berry curvature $\Omega_{12}$ multipoles under the 2D magnetic group $m'm'm$. 
    The subscript on $\partial_i$ and $\mathfrak{g}_{ij}$, $i,j=1,2$ indicates $k_x$ and $k_y$ for the rectangular lattices. 
    Each element in the table with column $\mathfrak{g}_{ij}$ and row $\partial_l$ should be understood as $\int_{\mathbf k} \sum_n f_n \partial_l \mathfrak{g}_{ij}$. A $0$ indicates the integral of the corresponding quantity over Brillouin zone must vanish.}
    \label{tab:constraint2}
\end{table}
\begin{table}[H]
    \centering
    \begin{tabular}{|c||c|c|c|c|}
    \hline
         $\int_{\mathbf k} \sum_n f_n$ &$\mathfrak{g}_{11}$ &$\mathfrak{g}_{22}$ &$\mathfrak{g}_{12}$ &$\Omega_{12}$ \\
    \hline
    \hline
         &$\mathfrak{g}_{22}$ &$\mathfrak{g}_{11}$ &$0$ &$0$ \\
    \hline
         $\partial_1$ &$0$ &$0$ &$0$ &$0$ \\
    \hline
         $\partial_2$ &$0$ &$0$ &$0$ &$0$ \\
    \hline
        $\partial_1\partial_1$ &$\partial_2\partial_2\mathfrak{g}_{22}$ &$\partial_2\partial_2\mathfrak{g}_{11}$ &$-\partial_2\partial_2\mathfrak{g}_{12}$ &$-\partial_2\partial_2 \Omega_{12}$ \\
    \hline
        $\partial_2\partial_2$ &$\partial_1\partial_1\mathfrak{g}_{22}$ &$\partial_1\partial_1\mathfrak{g}_{11}$ &$-\partial_1\partial_1\mathfrak{g}_{12}$ &$-\partial_1\partial_1 \Omega_{12}$\\
   \hline
        $\partial_1\partial_2$ &$-\partial_1\partial_2 \mathfrak{g}_{22}$ &$-\partial_1\partial_1 \mathfrak{g}_{12}$ &$\partial_1\partial_2 \mathfrak{g}_{12}$ &$\partial_1\partial_2 \Omega_{12}$ \\
    \hline
    \end{tabular}
    \caption{Constraint on quantum metric $\mathfrak{g}_{ij}$ multipoles and Berry curvature $\Omega_{12}$ multipoles under the 2D magnetic group $4'/m$. 
    } 
    \label{tab:constraint4}
\end{table}
\begin{table}[H]
    \centering
    \begin{tabular}{|c||c|c|c|c|}
    \hline
         $\int_{\mathbf k} \sum_n f_n$ &$\mathfrak{g}_{11}$ &$\mathfrak{g}_{22}$ &$\mathfrak{g}_{12}$ &$\Omega_{12}$ \\
    \hline
    \hline
         &$\mathfrak{g}_{22}$ &$\mathfrak{g}_{11}$ &$0$ &$0$ \\
    \hline
         $\partial_1$ &$0$ &$0$ &$0$ &$0$ \\
    \hline
         $\partial_2$ &$0$ &$0$ &$0$ &$0$ \\
    \hline
        $\partial_1\partial_1$ &$\partial_2\partial_2\mathfrak{g}_{22}$ &$\partial_2\partial_2\mathfrak{g}_{11}$ &$0$ &$-\partial_2\partial_2 \Omega_{12}$ \\
    \hline
        $\partial_2\partial_2$ &$\partial_1\partial_1\mathfrak{g}_{22}$ &$\partial_1\partial_1\mathfrak{g}_{11}$ &$0$ &$-\partial_1\partial_1 \Omega_{12}$\\
   \hline
        $\partial_1\partial_2$ &$0$ &$0$ &$\partial_1\partial_2 \mathfrak{g}_{12}$ &$0$ \\
    \hline
    \end{tabular}
    \caption{Constraint on quantum metric $\mathfrak{g}_{ij}$ and Berry curvature $\Omega_{12}$ multipoles under the 2D magnetic group $4'/mm'm$. }
    \label{tab:constraint41}
\end{table}
\begin{table}[H]
    \centering
    \begin{tabular}{|c||c|c|c|c|}
    \hline
         $\int_{\mathbf k} \sum_n f_n$ &$\mathfrak{g}_{11}$ &$\mathfrak{g}_{22}$ &$\mathfrak{g}_{12}$ &$\Omega_{12}$ \\
    \hline
    \hline
         &$\mathfrak{g}_{22}$ &$\mathfrak{g}_{11}$ &$0$ &$0$ \\
    \hline
         $\partial_1$ &$0$ &$0$ &$0$ &$0$ \\
    \hline
         $\partial_2$ &$0$ &$0$ &$0$ &$0$ \\
    \hline
        $\partial_1\partial_1$ &$\partial_2\partial_2\mathfrak{g}_{22}$ &$\partial_2\partial_2\mathfrak{g}_{11}$ &$0$ &$0$ \\
    \hline
        $\partial_2\partial_2$ &$\partial_1\partial_1\mathfrak{g}_{22}$ &$\partial_1\partial_1\mathfrak{g}_{11}$ &$0$ &$0$\\
   \hline
        $\partial_1\partial_2$ &$0$ &$0$ &$\partial_1\partial_2 \mathfrak{g}_{12}$ &$\partial_1\partial_2 \Omega_{12}$ \\
    \hline
    \end{tabular}
    \caption{Constraint on quantum metric $\mathfrak{g}_{ij}$ and Berry curvature $\Omega_{12}$ multipoles under the 2D magnetic group $4'/mmm'$. }
    \label{tab:constraint42}
\end{table}
\begin{table}[H]
    \centering
    \begin{tabular}{|c||c|c|c|c|}
    \hline
         $\int_{\mathbf k} \sum_n f_n$ &$\mathfrak{g}_{11}$ &$\mathfrak{g}_{22}$ &$\mathfrak{g}_{12}$ &$\Omega_{12}$ \\
    \hline
    \hline
         &$\mathfrak{g}_{22}$ &$\mathfrak{g}_{11}$ &$0$ &$0$ \\
    \hline
         $\partial_1$ &$0$ &$0$ &$0$ &$0$ \\
    \hline
         $\partial_2$ &$0$ &$0$ &$0$ &$0$ \\
    \hline
        $\partial_1\partial_1$ &$\partial_2\partial_2\mathfrak{g}_{22}$ &$\partial_2\partial_2\mathfrak{g}_{11}$ &$0$ &$0$ \\
    \hline
        $\partial_2\partial_2$ &$\partial_1\partial_1\mathfrak{g}_{22}$ &$\partial_1\partial_1\mathfrak{g}_{11}$ &$0$ &$0$\\
   \hline
        $\partial_1\partial_2$ &$0$ &$0$ &$\partial_1\partial_2 \mathfrak{g}_{12}$ &$0$ \\
    \hline
    \end{tabular}
    \caption{Constraint on quantum metric $\mathfrak{g}_{ij}$ and Berry curvature $\Omega_{12}$ multipoles under the 2D magnetic group $4/mmm$. }
    \label{tab:constraint43}
\end{table}

\begin{table}[H]
\centering
    \begin{tabular}{|c||c|c|c|c|}
    \hline
         $\int_{\mathbf k} \sum_n f_n$ &$\mathfrak{g}_{11}$ &$\mathfrak{g}_{22}$ &$\mathfrak{g}_{12}$ &$\Omega_{12}$ \\
    \hline
    \hline
         &$\mathfrak{g}_{22}$ &$\mathfrak{g}_{11}$ &$\frac12\mathfrak{g}_{11}$ &$0$ \\
    \hline
         $\partial_1$ &$0$ &$0$ &$0$ &$0$ \\
    \hline
         $\partial_2$ &$0$ &$0$ &$0$ &$0$ \\
    \hline
        $\partial_1\partial_1$ &$\partial_2\partial_2\mathfrak{g}_{22}$ &$\partial_1\partial_1\mathfrak{g}_{11}$ &$\frac12\partial_1\partial_1\mathfrak{g}_{11}$ &$0$ \\
    \hline
        $\partial_2\partial_2$ &$\partial_1\partial_1\mathfrak{g}_{22}$ &$\partial_2\partial_2\mathfrak{g}_{11}$ &$\frac12\partial_1\partial_1\mathfrak{g}_{11}$ &$0$ \\
   \hline
        $\partial_1\partial_2$ &$\frac12 \partial_1\partial_1\mathfrak{g}_{11}$ &$\frac12 \partial_1\partial_1\mathfrak{g}_{11}$ &$\frac14\partial_1\partial_1\mathfrak{g}_{11}$ &$0$\\
    \hline
    \end{tabular}
    \caption{Constraint on quantum metric $\mathfrak{g}_{ij}$ and Berry curvature $\Omega_{12}$ multipoles under the 2D magnetic groups $6/mmm$ and $6'/m'mm$. The subscript in $\partial_i$ and $\mathfrak{g}_{ij}$, $i,j=1,2$ indicates $k_1$ and $k_2$ for the hexagonal lattices. 
    To compute the longitudinal and transverse conductivities, one must convert BCQ and QMQ from this primitive basis to the orthogonal basis. 
    }
    \label{tab:constraint6}
\end{table}

\section{Non-linear conductivity in $d$-wave planar altermagnets}
In this appendix, we first show the symmetry allowed terms in the nonlinear conductivities for $d$-wave planar altermagnets. 
We then analytically compute the leading order contribution to each term in a representative effective model of a $d$-wave planar altermagnet with $4'/mm'm$ or $4'/mmm'$ symmetry group.

\subsection{Symmetry allowed nonlinear conductivity}
Applying the symmetry analysis to the third order responses in Eq.~(\ref{eqn:SM3rd}), we obtain the nonvanishing terms for the following conductivities
\begin{align}
    \sigma^{xxx;x} & = \tau^3 \frac{e^4}{\hbar^4}\int \sum_nf_n \partial_x\partial_x\partial_x\partial_x\epsilon + \tau \frac{e^4}{\hbar^2}\int \sum_nf_n  \partial_x\partial_x G^{xx} \\
    \sigma^{xxx;y} & = -\tau^2 \frac{e^4}{\hbar^3}\int \sum_nf_n \partial_x\partial_x\Omega^{xy} - \frac{e^4}{\hbar}\int \sum_nf_n \frac{2G^{xx}\Omega^{xy}}{\varepsilon_{n\bar n}}\\
    \sigma^{xxy;x} &= \tau^2 \frac{e^4}{\hbar^3} \int \sum_nf_n  \frac13 \partial_x\partial_x\Omega^{xy} + \frac{e^4}{\hbar}\int \sum_nf_n  \frac{2G^{xx}\Omega^{xy}}{3\varepsilon_{n\bar n}} = -\frac13 \sigma^{xxx;y}\\
    \sigma^{xyy;x} &= \tau^3 \frac{e^4}{\hbar^4} \int \sum_nf_n 
    \partial_x\partial_x\partial_y\partial_y\epsilon +\tau^2 \frac{e^4}{\hbar^3} \int \sum_nf_n  \frac23 \partial_x\partial_y\Omega^{xy} + \tau \frac{e^4}{\hbar^2} \int \sum_nf_n \frac13 \left( \partial_x\partial_x G^{yy} + 2 \partial_x\partial_y G^{xy} \right)
\end{align}
where we have used the abbreviation $x=k_x, y=k_y$.
For a $d_{x^2-y^2}$-wave altermagnet, $\int_k \partial_x\partial_y\Omega^{xy}=0$, $\int_k \partial_x\partial_x\Omega^{xy}\neq0$. However, after a $45^\circ$ rotation it is a $d_{xy}$-wave altermagnet with $\int_k \partial_x\partial_y\Omega^{xy}\neq0$ and $\int_k \partial_x\partial_x\Omega^{xy}=0$. The other terms are non-vanishing in general under rotations.

The other nonzero conductivities are related to the four listed conductivities by symmetry. For example, $\sigma^{yyx;y}=-\sigma^{xxy;x}$ due to $C_4{\cal T}$ symmetry.

\subsection{Analytical result near the Dirac point}
In this appendix, we analytically compute the nonlinear conductivities near each Dirac cone of Hamiltonian Eq.~(\ref{eqn:Ham_Toy}).
The Hamiltonian around a Dirac cone takes the following $k\cdot p$ form
\begin{align}
    H_{\Gamma} &\approx -\frac{J_1}{2}(k_x^2-k_y^2)\sigma_z + \frac{\lambda \sin(k_z/2)}{2}\left((k_x+k_y)\sigma_x+(k_y-k_x)\sigma_y\right) \\
    H_{M} &\approx \frac{J_1}{2}(k_x^2-k_y^2)\sigma_z + \frac{\lambda \sin(k_z/2)}{2}\left(-(k_x+k_y)\sigma_x+(k_y-k_x)\sigma_y\right)
\end{align}
The two Dirac cones share the same Berry curvature and quantum metric distribution. Therefore, we only need to study $H_\Gamma$ as an example. The eigenenergies are $E_\pm = \pm \sqrt{ \big(\frac{J_1}{2}(k_x^2-k_y^2)\big)^2+\lambda^2\sin^2(k_z/2)k_{\parallel}^2/2 }$ where $k_\parallel=\sqrt{k_x^2+k_y^2}$ is the in-plane momentum and $\pm$ label the up and bottom bands.  
For two level systems $H=\mathbf h \cdot \sigma$, the Berry curvature and quantum metric are given by~\cite{graf2021berry} 
\begin{align}
     \Omega_\pm^{ij} &=\mp \frac{1}{2|\mathbf{h}|^3} \mathbf{h} \cdot\left(\mathbf{h}^i \times \mathbf{h}^j\right) \\
     \mathfrak{g}_{ij,\pm} &=\frac{1}{4|\mathbf{h}|^2}\left[\mathbf{h}^i \cdot \mathbf{h}^j-\frac{\left(\mathbf{h} \cdot \mathbf{h}^i\right)\left(\mathbf{h} \cdot \mathbf{h}^j\right)}{|\mathbf{h}|^2}\right] 
\end{align}
where $\mathbf h^i = \partial_{k_i}\mathbf h$ and $\pm$ indicates the upper/lower band. We define the band normalized quantum metric $G_{ij,\pm} = \mathfrak{g}_{ij}/2E_\pm$. 
We further consider the limit $k\ll 1$. 
For $H_\Gamma$, the explicit forms of $\Omega$ and $G$ are
\begin{align}
    \Omega_\pm &= (\mp)\left(\frac{\lambda \sin(k_z/2)}{2}\right)^2\frac{J_1(k_x^2-k_y^2)}{2|E_\pm|^3} \\
    G^{xx}_{\pm} &= \left(\frac{\lambda \sin(k_z/2)}{2}\right)^2 \frac{J_1^2(k_x^4+6k_x^2k_y^2+k_y^4)+2k_y^2\lambda^2\sin^2(k_z/2)}{16|E_\pm|^5} \\
    G^{yy}_{\pm} &= \left(\frac{\lambda \sin(k_z/2)}{2}\right)^2 \frac{J_1^2(k_x^4+6k_x^2k_y^2+k_y^4)+2k_x^2\lambda^2\sin^2(k_z/2)}{16|E_\pm|^5} \\
    G^{xy}_{\pm} &= \left(\frac{\lambda \sin(k_z/2)}{2}\right)^2 \frac{k_xk_y \left( 2J_1^2(k_x^2+k_y^2)+\lambda^2\sin^2(k_z/2) \right)}{8|E_\pm|^5}
\end{align}
From these expressions, we can calculate the integrals $\int_{\mathbf k} \sum_n f_n \partial\partial\Omega_n $, $\int_{\mathbf k} \sum_n f_n \partial\partial G_n $ and $\int_{\mathbf k} \sum_n f_n \partial\partial\partial\partial \varepsilon_n $ which appear in the conductivities in Eq.~(\ref{eqn:SM3rd}). 
In the following for the sake of simplicity we work in the quasi-2d limit and set $k_z=\pi$. To understand how the BCQ, QMQ, Drude and AIC terms scale with $\mu$ around the Dirac points we take the limit of $k\ll \lambda/J$, where we can perform the analysis analytically. In this limit the eigen-energies are approximately $E_\pm = \pm \frac{\lambda }{\sqrt{2}} k_{\parallel}$.
Next, it is useful to use the polar basis $k_x=k_\parallel\cos\theta$, $k_y=k_\parallel\sin\theta$. 
Then the derivatives are $\partial_{k^x} = \cos\theta\partial_{k_\parallel}-(\sin\theta/k_\parallel)\partial_\theta$ and $\partial_{k^y} = \sin\theta\partial_{k_\parallel}+(\cos\theta/k_\parallel)\partial_\theta$. It follows that
\begin{align}
    \int_k \partial_x\partial_y\Omega^{xy} &\rightarrow 0\\
    \int_k \partial_x\partial_x\Omega^{xy} &\rightarrow  -\frac{\pi J_1}{8|\mu|} \left(1+O\big(\mu^2J_1^2/\lambda^4\big)\right) \label{eqn:SM_BCQ99}\\
    \int_k \partial_x\partial_xG^{xx} &\rightarrow \left(-\frac{5\pi \lambda^2}{64|\mu|^3} \right) \left(1+O\big(\mu^2J_1^2/\lambda^4\big)\right) \label{eqn:SM_xxxxG99} \\
    \int_k \partial_y\partial_yG^{xx} &\rightarrow \left(-\frac{7\pi \lambda^2}{64|\mu|^3}\right) \left(1+O\big(\mu^2J_1^2/\lambda^4\big)\right) \label{eqn:SM_yyxxG99}\\
    \int_k \partial_x\partial_yG^{xy} &\rightarrow  \left(-\frac{\pi \lambda^2}{64|\mu|^3} \right) \left(1+O\big(\mu^2J_1^2/\lambda^4\big)\right) \label{eqn:SM_xyxyG99}
\end{align}
and 
\begin{align}
    \text{AIC} \propto \int_k \frac{G^{xx}\Omega^{xy} }{2\epsilon}&\rightarrow -\frac{\pi J}{384 |\mu|^3} \left(1+O\big(\mu^2J_1^2/\lambda^4\big)\right) \label{eqn:SM_AIC99}\\
    \int_k \partial_x\partial_x \epsilon &\rightarrow -\pi |\mu| \left(1+O\big(\mu^2J_1^2/\lambda^4\big)\right) \label{eqn:SM_xxD99}\\
    \int_k \partial_x\partial_x\partial_x\partial_x \epsilon &\rightarrow -\frac{3\pi \lambda^2}{8|\mu|} \left(1+O\big(\mu^2J_1^2/\lambda^4\big)\right) \label{eqn:SM_xxxxD99}\\
    \int_k \partial_x\partial_x\partial_y\partial_y \epsilon &\rightarrow -\frac{\pi \lambda^2}{8|\mu|} \left(1+O\big(\mu^2J_1^2/\lambda^4\big)\right) \label{eqn:SM_xxyyD99}
\end{align}

The leading order terms show that the quantum metric quadrupole terms (Eqs.~(\ref{eqn:SM_xxxxG99}), (\ref{eqn:SM_yyxxG99}) and (\ref{eqn:SM_xyxyG99})) are greater than the Drude terms (Eqs.~(\ref{eqn:SM_xxxxD99}) and (\ref{eqn:SM_xxyyD99})) when $\mu\rightarrow 0$.
The AIC term enhances to contribution to the conductivity from the Berry curvature quadrupole.
Comparing Eq.~(\ref{eqn:SM_BCQ99}) and Eq.~(\ref{eqn:SM_AIC99}),
the AIC term dominates.

\section{Spin-group protected altermagnetic nodal lines}
As we discussed in the main text, for altermagnets there is a sharp contribution to nonlinear response originated from the gap induced by SOC in spin-group protected nodal lines. Here, we explicitly show the spin-group origin of these nodal lines for the example of $d$-wave altermagnets in $4'/mm'm$ group. 
The $d$-wave altermagnets with SOC are in the magnetic space group $4'/mm'm$. The symmetries include $C_{4z}{\cal T}$, inversion $\cal I$, mirror $M_{110}$, and $M_{100}{\cal T}$. The Dirac points at $\Gamma$ and $M$ are protected by symmetries of the little group $4'/mm'm$.~\cite{bradley2010mathematical}. 

When SOC is absent, a planar $d$-wave altermagnet is in the spin group ${}^24/{}^1m{}^1m{}^2m \times {\mathbb Z}_2^{[C_{2x}||E]{\cal T} } \times U(1)_z$. The anti-unitary symmetry $[C_{2x}||E]{\cal T}$ and the continuous rotation symmetry about the $z$-axis in the spin space $U(1)_z$ are consequences of having collinear spin ordering with the Neel vector in the $z$-direction.
The unitary symmetries include $[C_{2x}||C_4]$, $[E||{\cal I}]$, $[E||M_{100}]$ and $[C_{2x}||M_{110}]$~\cite{litvin1974spin}. Note that conjugating by a rotation in $U(1)_z$ also implies the symmetry $[C_{2{\mathbf n}}||C_4]$ for generic unit vector $\mathbf n$ in the $k_x$-$k_y$ plane.

We now describe how the spin group symmetries protect nodal line crossings along $k_x=\pm k_y$ for $d_{x^2-y^2}$-wave altermagnets. 
$U(1)_z$ is an internal symmetry that leaves each momentum $\mathbf k$ invariant. Thus, wavefunctions at each $\mathbf k$ point must belong to one of its two irreducible representations, spin up or spin down (we restrict ourselves to spin-$1/2$ representations). When a spin up band crosses a spin down band, the crossing is protected because no symmetry-preserving term can couple the two irreps. 
Since $U(1)_z$ is a symmetry for any system with collinear spins, such crossings are protected by symmetry for all altermagnets. 
We now prove such crossings are required in the $d$-wave altermagnets:
since $[C_{2x}||C_4]$ symmetry maps spin up at $\mathbf k$ to spin down at $R_4 \mathbf k$, the spin up/down bands must cross at intermediate points.
Thus, any path from a point $\mathbf{k}$ to its rotated counterpart $R_4\mathbf{k}$ must contain a band crossing.

Spin group mirror symmetries can pin the positions of these crossings.
Specifically, $[C_{2x}||M_{110}]$ leaves the lines $k_x = \pm k_y$ invariant. Since $[C_{2x}||M_{110}]$ 
anti-commutes with $[C_{2z}||E]$ -- which acts trivially in real and momentum space -- the group generated by these elements only has two-dimensional irreps.
Thus, bands are required to come in degenerate pairs, i.e., nodal lines, along the $k_x = \pm k_y$ lines.
This corresponds to our $d_{x^2-y^2}$ wave altermagnet, which describes RuO$_2$.

On the other hand, if $[C_{2x}||M_{100}]$ is a symmetry, 
then a nodal line is protected along $k_x=0$ or $k_y=0$ by the same logic. This corresponds to a $d_{xy}$-wave altermagnet. 

SOC breaks the spin group $P{}^24/{}^1m{}^1m{}^2m \times {\mathbb Z}_2^{[C_{2x}||E]{\cal T} } \times U(1)_z$ down to the magnetic group $4'/mm'm$, gapping
the nodal lines and creating anti-crossings.

\section{Ruthenium Oxide: bulk and thin film models}

\subsection{Bulk Hamiltonian}
\begin{figure}[H]
    \centering
    \includegraphics[width=0.7\linewidth]{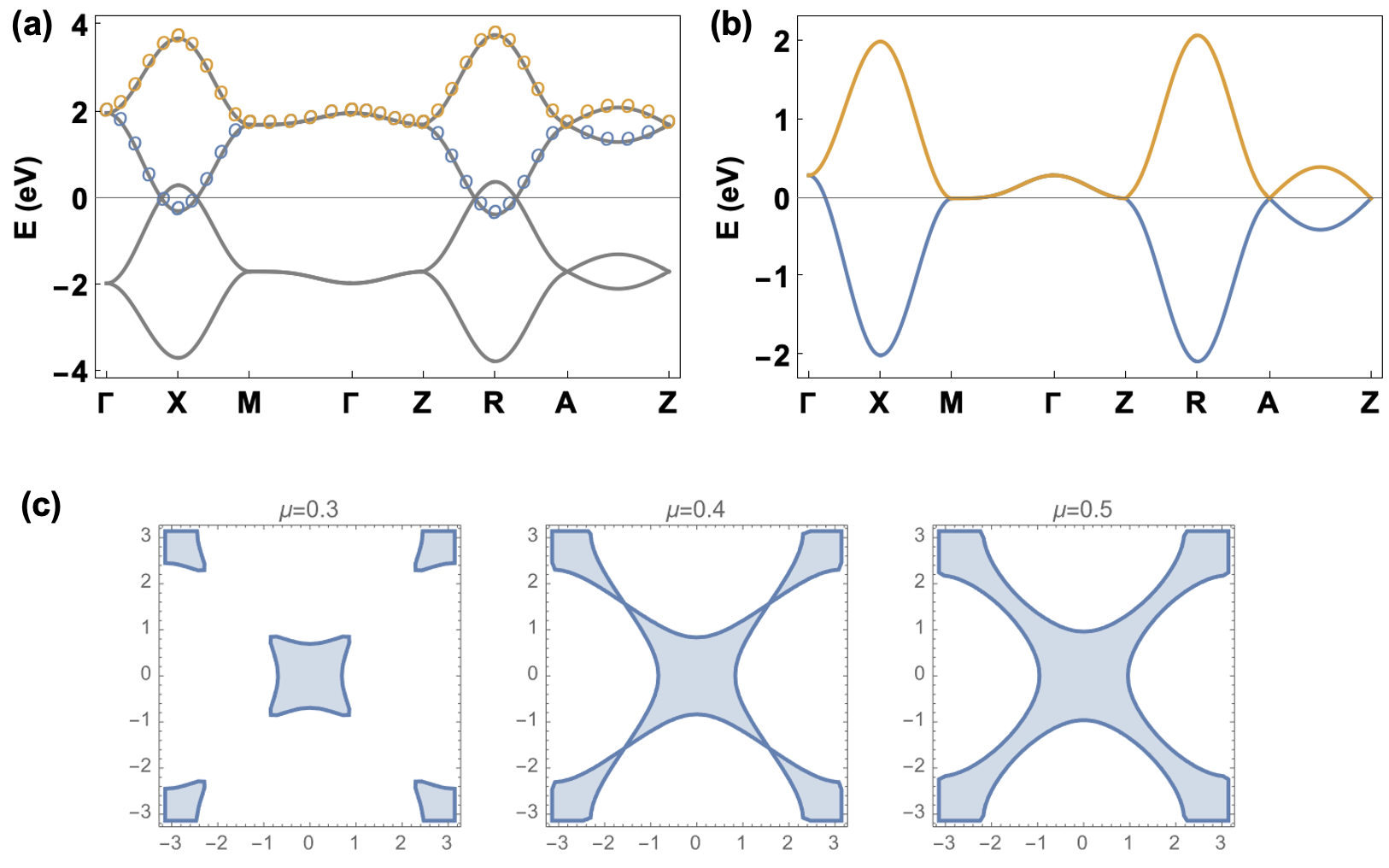}
    \caption{(a) Spectrum of the three dimensional RuO$_2$ four band model (gray) with parameters $J_0=1.7$eV, $J_1=1$eV, $t=1$eV, $\lambda=0.4$eV. The spectrum of the effective two band model for the upper two bands are plotted in colored circles, which agrees with the four band model. 
    (b) Spectrum of the effective two band model corresponding to colored circles in (a) using the same parameters with an overall shift. (c) Fermi sea of the two band model at $k_z=\pi$ with $\mu=0.3,0.4,0.5$ eV. The van Hove singularity is at $\mu_{vH}=\lambda=0.4$eV. }
    \label{fig:SM1}
\end{figure}

We now construct a two band toy model to describe the physics of RuO$_2$, starting from a four band model describing two spins and two orbitals at the $C_{4z,2}{\cal T}$ related $A$ and $B$ sublattices (located at $(\frac12,0,0)$ and $(0,\frac12,\frac12)$ respectively)~\cite{vsmejkal2020crystal}: 

\begin{align}
H_0 =&  t_0\cos(\frac{k_x}2)\cos(\frac{k_y}2)\cos(\frac{k_z}{2})\sigma_0\tau_x + J_0 \sigma_z\tau_z + J_1 (\cos k_x -\cos k_y)\sigma_z\tau_0 \cr
&+\lambda(\sin(\frac{k_x+k_y}{2})\sigma_x+\sin(\frac{k_y-k_x}{2})\sigma_y)\sin(\frac{k_z}{2})\tau_x 
\label{eqn:H0}
\end{align}
where $t_0$ describes nearest neighbor hopping, $J_0$ comes from the band separating zero-th order altermagnet term.
$J_0$ and $J_1$ are the symmetry allowed onsite and next-nearest hopping altermagnetic terms that split the spin degeneracy (as marked in Fig~\ref{fig:SM1}(a)), $\sigma$ and $\tau$ are Pauli matrices in the spin and orbital spaces.
When $\lambda = 0$, the Hamiltonian is invariant under the spin group ${}^24/{}^1m{}^1m{}^2m$. In the presence of SOC, it is described by the magnetic group $4'/mm'm$.

When $J_0$ dominates the Hamiltonian, the four bands split into two groups, each describing an atomic limit with spins/orbitals $|A\uparrow\rangle$, $|B\downarrow\rangle$ or $|A\downarrow\rangle$, $|B\uparrow\rangle$. The effective two-band Hamiltonian takes the form
\begin{align}
\label{eqn:Hatkz}
    H_{s=\pm} &= s E_0(k)+J_1 (\cos k_x -\cos k_y)\sigma_z +\lambda \left(\sin \left( \frac{k_x+k_y}{2}\right)\sigma_x+\sin \left(\frac{k_y-k_x}{2} \right)\sigma_y \right) \sin \left(\frac{k_z}{2} \right)  \\
\label{eqn:Hatkz_J0}
    E_0(k) & = J_0\left(1+\frac12 \frac{t_0^2}{J_0^2}\cos^2\frac{k_x}{2}\cos^2\frac{k_y}{2}\cos^2\frac{k_z}{2} \right)
\end{align}
where $\sigma$'s here are the Pauli matrices that describe the mixed degrees of freedom, $|A\uparrow\rangle$ and $|B\downarrow\rangle$.
The effective Hamiltonian takes the same form as our toy model in Eq.~(\ref{eqn:Ham_Toy}), plus the diagonal term $E_0(k)$, which shifts the Dirac points at $\Gamma$ and $M$ in energy.
The spectrum of the four band Hamiltonian Eq.~(\ref{eqn:H0}) is plotted in gray in Fig.~\ref{fig:SM1}(a) with parameters $J_0=1.7$eV, $J_1=1$eV, $t=1$eV. The spectrum of effective two band model for the top bands Eq.~(\ref{eqn:Hatkz}) overlays on the same plot. 
In Fig.~\ref{fig:SM1}(b) we shift the effective two bands by $-J_0$. This two band model at $k_z=\pi$ is exactly the same as our toy model Eq.~(\ref{eqn:Ham_Toy}). In Fig.~\ref{fig:SM1}(c) the Fermi sea in $k_z=\pi$ and $\mu=0.3$, $0.4$, $0.5$eV are shown. The van Hove singularity is at $\mu_{vH}=\lambda=0.4$eV in this $k_z=\pi$ plane. 

\subsection{Thin film model}
For the two-band Hamiltonian Eq.~(\ref{eqn:Hatkz}),
we consider boundary conditions
periodic in $x$ and $y$ but finite in the $z$-direction with $0\leq z\leq d$ and hard wall boundary conditions.
The Hamiltonian for this thin film can be approximated by replacing $k_z$ with $-i\partial/\partial z$.
Consider the ansatz $\psi(k_x,k_y,z) = e^{\alpha(k_x,k_y) z} \chi(k_x,k_y)$ where $ \chi(k_x,k_y)$ is a spinor. 
The Hamiltonian is symmetric under $\alpha \rightarrow 2\pi i-\alpha$. Therefore, the general solution of the wavefunction with the hard wall boundary condition takes the form $\Psi_n(k_x,k_y,z) = C_n~\sin(k_n z) \chi(k_x,k_y)$, 
where $k_n = n\pi /d$, $n=1,\dots,d-1$ and $C_n$ is a normalizing constant. The thin film Hamiltonian acting on this wavefunction is
\begin{align}\label{thinfilm}
     H_n(k_x,k_y) &= t_n \cos^2\frac{k_x}{2}\cos^2\frac{k_y}{2}+J_1 (\cos k_x -\cos k_y)\sigma_z +\lambda_n \left(\sin \left( \frac{k_x+k_y}{2}\right)\sigma_x+\sin \left(\frac{k_y-k_x}{2} \right)\sigma_y \right) 
\end{align}
where $t_n=t^2\cos^2(k_n/2)/{2J_0}$ and $\lambda_n=\lambda\sin(k_n/2)$.

We used Eq.~\eqref{thinfilm} to compute the nonlinear conductivity of RuO$_2$ in Fig.~4 of the main text. 


\end{document}